\documentclass[iop]{emulateapj}

\shorttitle{Tracing the dark energy history with GRBs}

\graphicspath{{./}{figures/}}

\usepackage{amssymb}
\usepackage{amsmath}
\usepackage{graphicx}
\usepackage{natbib}
\usepackage{multirow}
\usepackage{hhline}
\usepackage{txfonts}
\usepackage{color}
\usepackage{epstopdf}
\usepackage{microtype}
\usepackage[colorlinks=true,citecolor=blue,linkcolor=blue]{hyperref}
\usepackage[flushleft]{threeparttable}
\usepackage{booktabs}

\defcitealias{Izzo2015}{I15}
\defcitealias{RuffiniMuccino2014}{R14}
\defcitealias{Bernardini2012}{B12}
\defcitealias{Margutti2013}{M13}
\defcitealias{Planck2018}{P20}
\defcitealias{2019ApJ...876...85R}{R19}

\bibpunct{(}{)}{;}{a}{}{,}

\begin{document}

\title{Tracing dark energy history with gamma ray bursts}

\author{M.~Muccino\altaffilmark{1,2},
L.~Izzo\altaffilmark{3},
O.~Luongo\altaffilmark{2,4,5},
K.~Boshkayev\altaffilmark{2,6},
L.~Amati\altaffilmark{7},
M.~Della~Valle\altaffilmark{8},
G.~B.~Pisani,
E.~Zaninoni}

\email{marco.muccino@lnf.infn.it, luca.izzo@nbi.ku.dk}
\email{orlando.luongo@unicam.it}

\altaffiltext{1}{INFN, Laboratori Nazionali di Frascati, Via Enrico Fermi, 54, 00044, Frascati (RM), Italy}
\altaffiltext{2}{NNLOT, Al-Farabi Kazakh National University, Al-Farabi av. 71, 050040 Almaty, Kazakhstan}
\altaffiltext{3}{DARK, Niels Bohr Institute, University of Copenhagen, Lyngbyvej 2, DK-2100 Copenhagen, Denmark}
\altaffiltext{4}{Dipartimento di Matematica, Universit\`a di Pisa, Largo B. Pontecorvo 5, Pisa, 56127, Italy}
\altaffiltext{5}{Divisione di Fisica, Universit\`a di Camerino, Via Madonna delle Carceri, 9, 62032, Italy}
\altaffiltext{6}{Department of Physics, Nazarbayev University, 010000 Nur-Sultan, Kazakhstan}
\altaffiltext{7}{INAF, Istituto di Astrofisica Spaziale e Fisica Cosmica, Bologna, Via Gobetti 101, Bologna, 40129, Italy}
\altaffiltext{8}{INAF, Osservatorio Astronomico di Capodimonte, Salita Moiariello 16,  Napoli, 80131, Italy}

\begin{abstract}
Observations of gamma-ray bursts up to $z\sim 9$ are best suited to study the possible evolution of the Universe equation of state at intermediate redshifts. We apply the Combo-relation to a sample of 174 gamma ray bursts to investigate possible evidence of evolving dark energy parameter $w(z)$.
We first build a gamma ray burst Hubble's diagram and then we estimate the set ($\Omega_m$, $\Omega_{\Lambda}$) in the framework of flat and non-flat $\Lambda$CDM paradigm. We then get bounds over the $w$CDM model, where $w$ is thought to evolve with redshift, adopting two priors over the Hubble constant in tension at $4.4$-$\sigma$, i.e.  $H_0=(67.4\pm0.5)$~km/s/Mpc and $H_0=(74.03\pm1.42)$~km/s/Mpc. We show our new sample provides tighter constraints on $\Omega_m$ since at $z\leq1.2$ we see that $w(z)$ agrees within 1$\sigma$ with the standard value $w=-1$. The situation is the opposite at larger $z$, where gamma ray bursts better fix $w(z)$ that seems to deviate from $w=-1$ at $2$--$\sigma$ and $4$--$\sigma$ level, depending on the redshift bins. In particular, we investigate the $w(z)$ evolution through a piecewise formulation over seven  redshift intervals. From our fitting procedure we show that at $z\geq 1.2$ the case $w<-1$ cannot be fully excluded, indicating that dark energy's influence is not negligible at larger $z$. We confirm the Combo relation as a powerful tool to investigate cosmological evolution of dark energy. Future space missions will significantly enrich the gamma ray burst database even at smaller redshifts, improving \emph{de facto} the results discussed in this paper.
\end{abstract} 

\keywords{Standard candles, Gamma-ray burst, dark energy}




\section{Introduction}\label{sec:1}

In the cosmological concordance model, the Universe is approximated by $\sim30\%$ of baryonic and cold dark matter and by $\sim70\%$ of an exotic form of constant \emph{dark energy} \citep[DE, see e.g.,][hereafter P20]{Planck2018}. In particular to speed up the Universe today, DE counterbalances the action of gravity through  a negative pressure. Recent observations at small redshifts favor a cosmological constant contribution, $\Lambda$, to evolving DE, albeit  recent observations provided by \citet{Planck2018} seem to show unexpected tensions among observables, not yet understood within the concordance paradigm, dubbed the $\Lambda$CDM model. In the latter, the corresponding equation of state (EoS) turns out to be \emph{exactly} $w\simeq -1$ (\citetalias{Planck2018}; \citealt{Riess2007}). Any deviations from $w=-1$ would lead to more complicated versions of DE fluids or to
extended and/or modified theories of gravity
still objects of debate \citep{Bronstein1933,Tsujikawa2013,Sahni2014,Ding2015,orlando1}.

Several methods have been proposed to investigate a possible evolution of $w(z)$ with the redshift, mainly involving supernovae (SNe) Ia \citep{Phillips1993,1998Natur.391...51P,Perlmutter1999,Riess1998,Schmidt1998} and/or standard candles and rulers. However, the shortage of data at high redshifts ($z\geq1$) inevitably brings large uncertainties over $w$ measurements \citep{Union2.1}, providing an unvoidable degeneracy problem among cosmological models. For example, the simplest $\Lambda$CDM generalization is based on a first-order approximation, $w(z) = w_0 + w_a z/(1+z)$ \citep{Chevallier2001,Linder2003} fully degenerating with several other  redshift-binned parameterizations within the sphere $z<1$ \citep[see, e.g.,][]{King2014}.

\begin{figure}
\centering
\includegraphics[width=\hsize,clip]{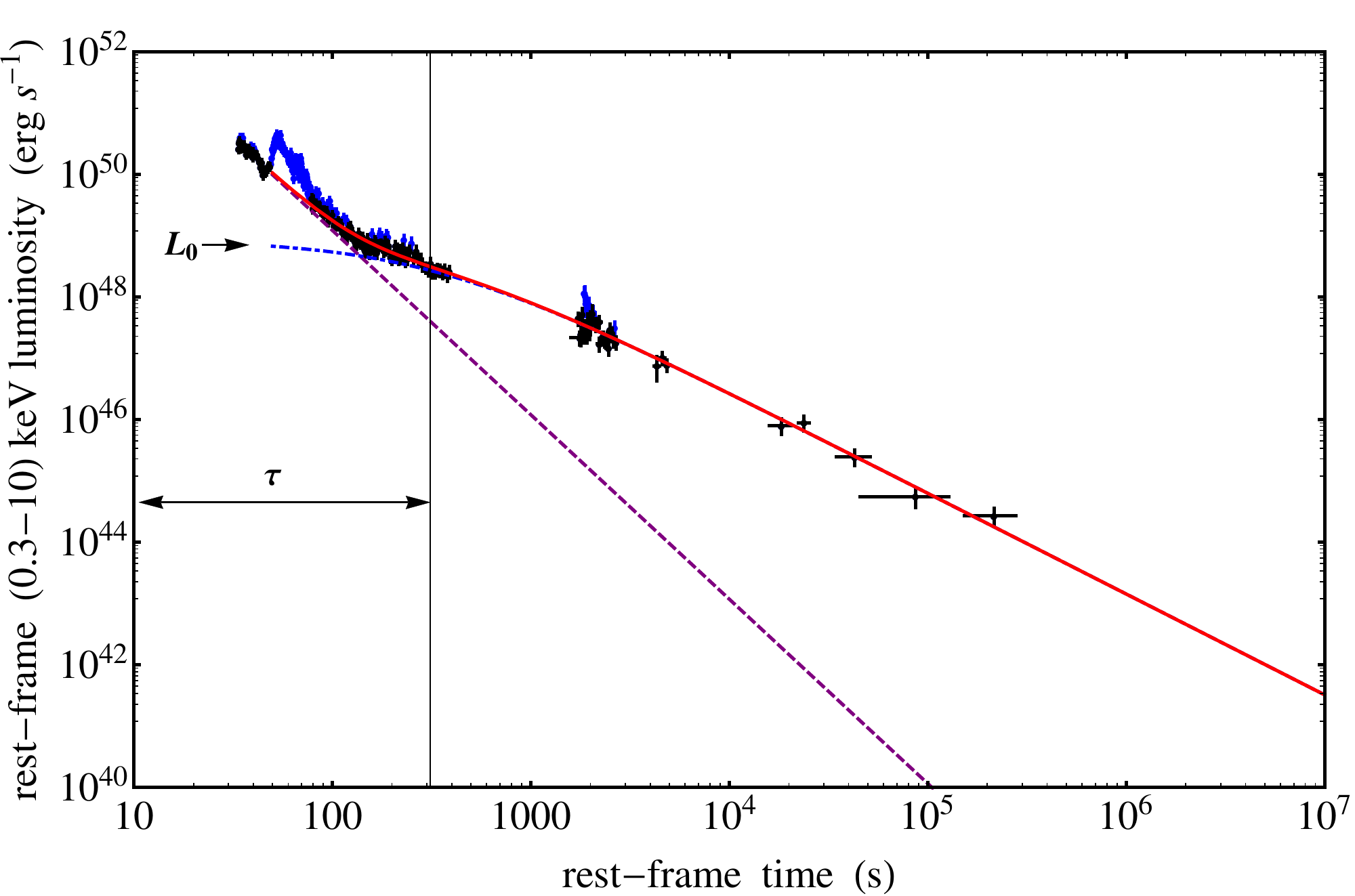}
\caption{The rest-frame $0.3$--$10$~keV light curve of GRB 060418A. The total fit (red curve) is composed of a steep decay (dashed purple power-law) and a plateau + late power-law decay (dot-dashed blue curve). The black dots are the data filtered by the flares (blue dots). The vertical black line indicates $\tau$; an arrow marks the luminosity $L_0$.}
\label{fig:MDV}
\end{figure}

Gamma ray bursts (GRBs) have the advantage over other cosmological probes and rulers to homogeneously cover a large interval of redshift up to $z \approx 9$ \citep{Cucchiara2011,2012ApJ...749...68S} and their redshift distribution peaks at $z\sim2$--$2.5$ \citep{Coward2013}, where the farthest SN Ia has been detected \citep{Rodney2015}. In the last decades, the analysis of larger samples has led to various phenomenological correlations between GRB photometric and spectroscopic properties,  suggesting  possible cosmological applications \citep[see e.g.][and references therein]{Schaefer2007,Amati2008,CapozzielloIzzo2008,Dainotti2008,Izzo2009,AmatiDellaValle2013,Wei2014}. Recently \citet{Izzo2015} (hereafter I15) developed a method for measuring the cosmological parameters from a sample of $60$ GRBs, which minimizes the cosmological ``circularity'' problem affecting the computation of the luminosity distances in all GRB correlations\footnote{For alternative techniques see e.g. \citep{orlando2,orlando3,orlando4}.}. This technique is based on the so-called \textit{Combo relation},\footnote{The name \textit{Combo} states for \emph{combination} since it combines $E_{\rm p,i}$--$E_{\rm\gamma,iso}$ \citep{Amati2008} and $E_{\rm\gamma,iso}$--$E_{\rm X,iso}$--$E_{\rm p,i}$ relations \citep{Bernardini2012}.} which is characterized by a small data scatter and involves prompt and afterglow GRB parameters \citep{Bernardini2012,Margutti2012}. The ``candle'' is provided by the plateau X-ray afterglow luminosity $L_0$, which is related to its rest-frame duration $\tau$, to the late power-law decay index $\alpha$ of the X-ray afterglow luminosity, and to the rest-frame peak energy of the GRB prompt emission $E_{\rm p,i}$ (see Fig.~\ref{fig:MDV}). The existence of this relation has been later on confirmed by \citet{2016ApJ...825L..20D} where a similar relation based on the use of the peak luminosity has been derived.

In this work, adopting an extended GRB sample, characterized by a complete data set in gamma- and X-rays, we fix limits over DE's evolution. Our general strategy remarks the same of \citetalias{Izzo2015} and aims at getting  novel bounds over the whole matter content, i.e. $\Omega_m$. Afterwards, we fix the cosmological constant density, $\Omega_\Lambda$, testing the validity of the $\Lambda$CDM model. Our findings adopt the estimated values of $L_0$ as distance indicator and can be generalized to any DE scenarios. To this end, we constrain the evolution of $w(z)$ through a piecewise formulation over the GRB redshift intervals.  Theoretical discussions of our results are prompted, leading to an overall agreement with the $\Lambda$CDM model at small and intermediate redshifts. Outside our time, slight deviations from $w=-1$ are not excluded \emph{a priori}. Precisely, from our new GRB sample we get tighter constraints on DE's evolution, showing that at $z\leq 1.2$ DE turns out to be compatible with a cosmological constant, but beyond $z\simeq1.2$ it does not since the case $w<-1$ cannot be fully excluded. These deviations are however severely affected by uncertainties and so further space missions, e.g. {\it Space Variable Objects Monitor} \citep[SVOM,][]{SVOM}, {\it enhanced X-ray Timing and Polarimetry} \citep[eXTP,][]{eXTP}, and {\it Transient High-Energy Sky and Early Universe Surveyor} \citep[THESEUS,][]{2018AdSpR..62..191A}, will significantly enrich the GRBs database to remove the ambiguity over the form of $w(z)$ as we discuss later in the manuscript.

The paper is structured as follows. In Section~\ref{sec:2}, we extend the sample of $60$ GRBs in \citetalias{Izzo2015} with $114$ additional GRBs characterized by a complete data set in gamma- and X-rays up to December 2018. In Section~\ref{sec:3}, we provide new constraints on the cosmological parameters $\Omega_m$ and $\Omega_\Lambda$ using the estimated values of $L_0$ as distance indicator. In particular, in Section~\ref{sec:3.4}, we constrain the evolution of $w(z)$ by considering a piecewise formulation over the GRB redshift interval. The results of our analysis are then discussed in Section~\ref{sec:7},
In Section~\ref{sec:8} we draw our conclusions.

\section{The updated GRB sample}\label{sec:2}

\subsection{Brief summary on the relation and the updated sample}\label{sec:2.1}

The Combo relation writes as \citepalias[see][for details]{Izzo2015}
\begin{equation}
\label{eq:no2}
\log \left(\frac{L_0}{{\rm erg/s}}\right) = \log \left(\frac{A}{{\rm erg/s}}\right) + \gamma \log \left(\frac{E_{\rm p,i}}{{\rm keV}}\right) - \log \left(\frac{\tau/{\rm s}}{|1+\alpha|}\right)\ ,
\end{equation}
where $\gamma$ is the slope and $A$ the normalization.
For each GRB, the rest-frame peak energy $E_{\rm p,i}$ is inferred from the $\nu F(\nu)$ GRB spectrum, while $L_0$, $\tau$, and $\alpha$ by fitting the rest-frame $0.3$--$10$~keV flare-filtered afterglow luminosity light curves with the function $L(t)=(1+t/\tau)^\alpha$ introduced in \citet{RuffiniMuccino2014} (hereafter R14, see, e.g., Fig.~\ref{fig:MDV}).\footnote{The flare-filtered luminosity light curves are iteratively fitted with the function in \citetalias{RuffiniMuccino2014}: at every iteration, data points with the largest positive residual are discarded, until a final fit with a p-value $>0.3$ is obtained.}

\begin{figure}
\centering
\includegraphics[width=\hsize,clip]{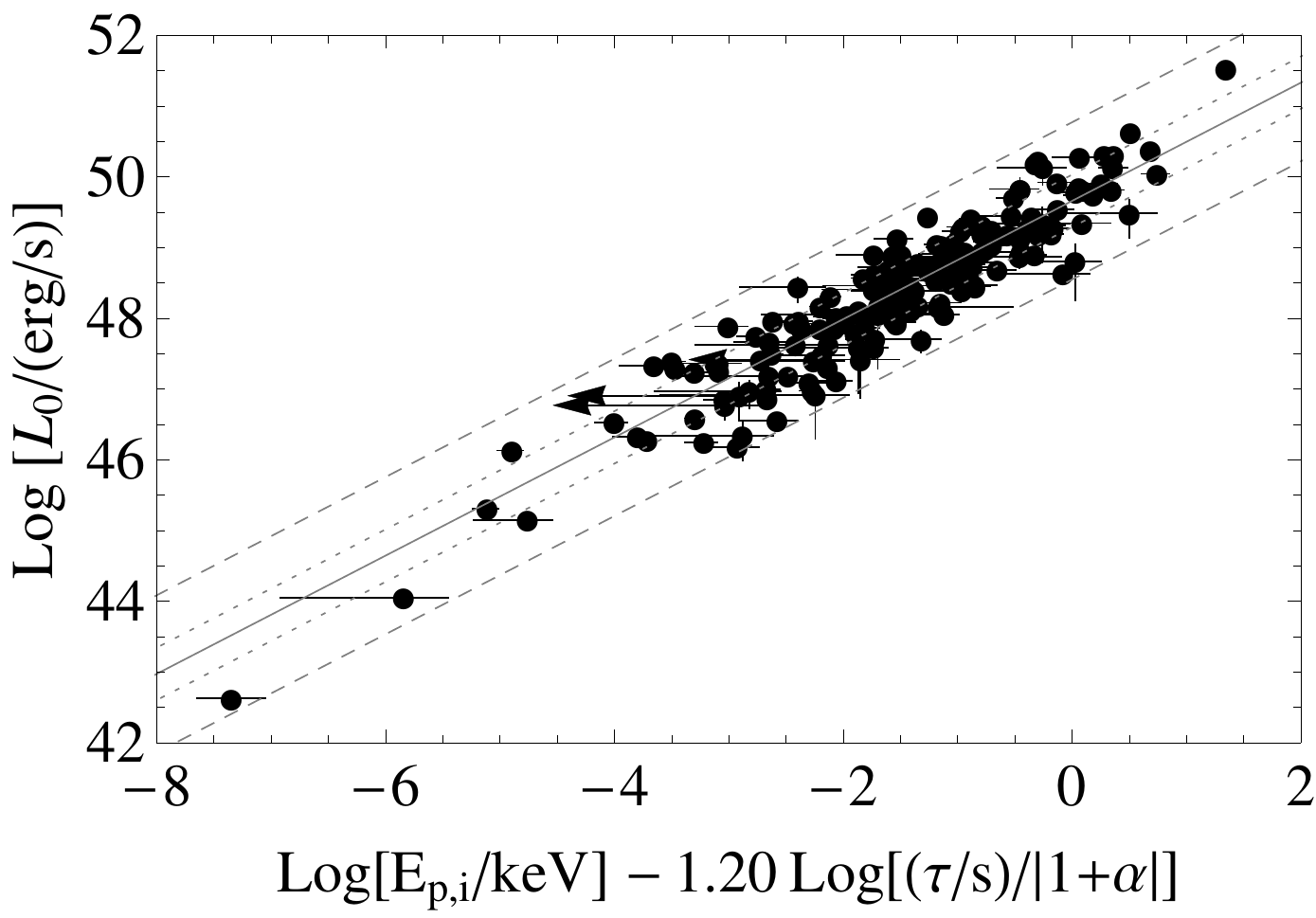}
\caption{The Combo relation from the sample of $174$ GRBs (black circles) in the flat $\Lambda$CDM model with indicative values $H_0=71$ km/s/Mpc, $\Omega_m=0.27$, and $\Omega_\Lambda=0.73$. The best-fit (solid gray line) and the $1$- and $3$--$\sigma$ of extra-scatter (dotted and dashed gray lines) are also displayed.}
\label{fig:no}
\end{figure}

From our sample we exclude light curves with $\alpha>-1$, which may have a change in slope beyond the XRT time coverage and/or be polluted by a late flaring activity.
We obtain a sample of {\bf$174$} GRBs, of which $60$ from \citetalias{Izzo2015} and {\bf$114$} introduced in this work (see Table~\ref{tab:no}). The Combo relation within the flat $\Lambda$CDM model is displayed in Fig.~\ref{fig:no}. Following \citet{Dago2005}, we obtain the best-fit parameters: $\log[A/({\rm erg/s})]=49.66\pm0.20$, $\gamma=0.84\pm0.08$, and an extra-scatter $\sigma_{\rm ex}=0.37\pm0.02$. The corresponding Spearman's coefficient is $\rho_s=0.92$ and the p-value from the two-sided Student's $t$-distribution is $p=1.17\times10^{-48}$.

\subsection{The calibration}\label{sec:2.2}

The Combo relation is calibrated through a two steps method minimizing the use of SNe Ia \citepalias{Izzo2015}. This method is based on:
\begin{itemize}
\item[(1)]{the determination of $\gamma$ and $\sigma_{\rm ex}$ from various, small sub-samples of GRBs lying almost at the same redshift;}
\item[(2)]{the determination of $A$ from SNe Ia located at the same redshift of the nearest GRBs of the sample.}
\end{itemize}
The use of SNe Ia in the step (2) is limited to the lowest redshift where the effect of the cosmology on the distance modulus of the calibrating SNe Ia is negligible (see Fig.~$3$ in \citetalias{Izzo2015}).

\subsubsection{The determination of $\gamma$ and $\sigma_{ex}$}\label{sec:2.2.1}

\begin{figure}
\centering
\includegraphics[width=\hsize,clip]{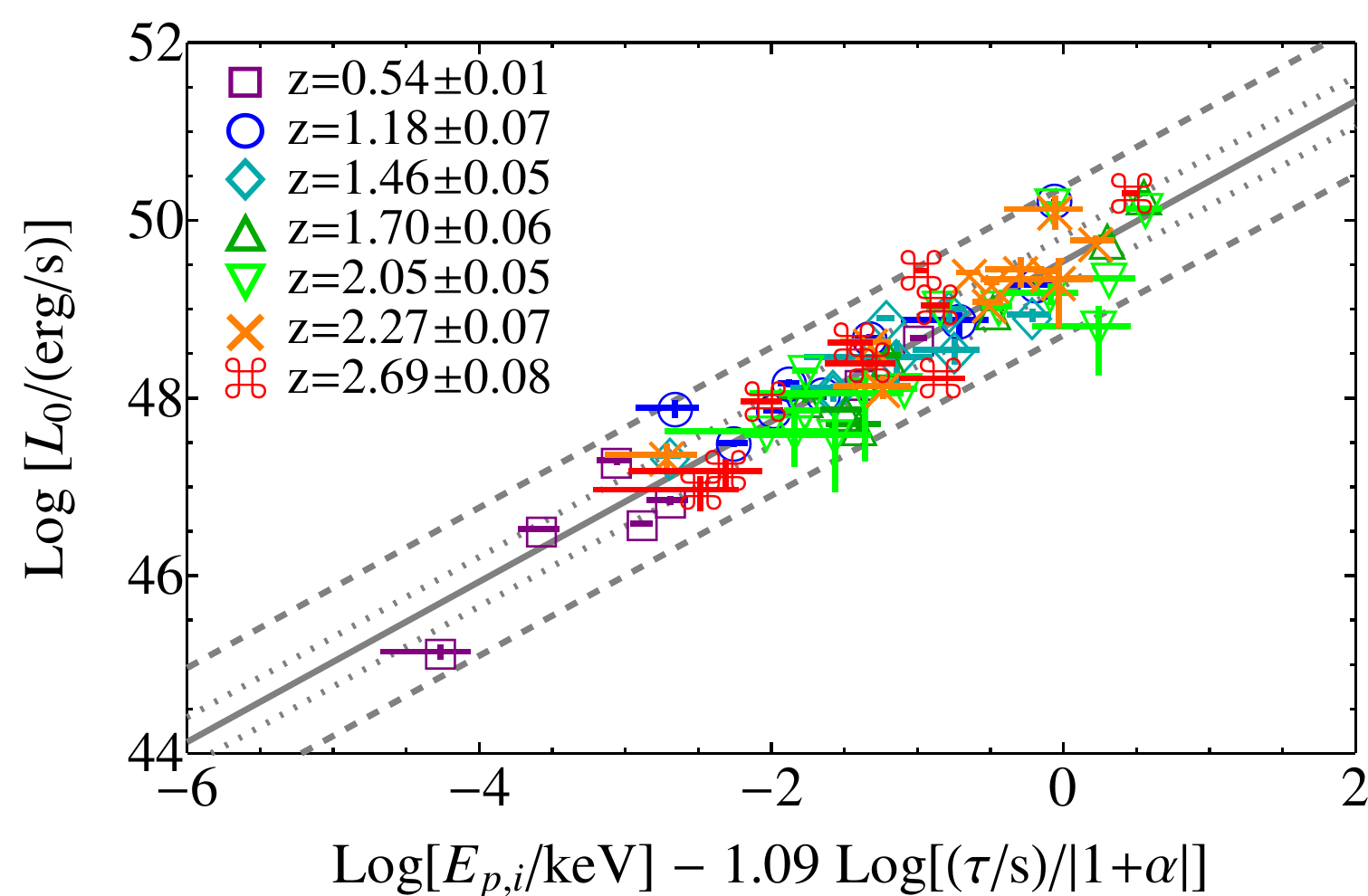}
\caption{The Combo sub-samples (see legend). The solid gray best-fit line and the dotted (dashed) gray lines at $1$--$\sigma$ ($3$--$\sigma$) are obtained from the calibrated values of $\gamma$ and $\sigma_{\rm ex}$ as inferred in Sec.~\ref{sec:2.2.1}. The normalization $A$ is constrained to the value determined in Sec.~\ref{sec:2.2.2}.}
\label{fig:no2}
\end{figure}

\begin{table}
\centering
\caption{Best-fit parameters of the seven sub-samples at average redshift $\langle z\rangle$: the slope $\gamma_z$, the normalization $A_{\rm z}$, and the extrascatter $\sigma_{\rm ex,z}$ are shown.}
\begin{threeparttable}
\begin{tabular}{c|ccc}
\hline\hline
$\langle z\rangle$  &   $\gamma_z$       &  $\log \left(A_{\rm z}/{\rm cm}^2/{\rm s}\right)$    & $\sigma_{\rm ex,z}$ \\
\hline
\hline
$0.54\pm0.01$	&	$0.81\pm0.49$	&	$-7.38\pm1.08$	                                    &	$0.29\pm0.10$	\\
$1.18\pm0.07$	&	$0.83\pm0.31$	&	$-7.93\pm0.77$	                                    &	$0.26\pm0.08$	\\
$1.46\pm0.05$	&	$0.80\pm0.32$	&	$-8.29\pm0.79$	                                    &	$0.26\pm0.08$	\\
$1.70\pm0.06$	&	$0.94\pm0.32$	&	$-8.92\pm0.84$	                                    &	$0.19\pm0.08$	\\
$2.05\pm0.05$	&	$1.05\pm0.32$	&	$-9.44\pm0.90$							            & $0.34\pm0.08$ \\
$2.27\pm0.07$	&	$0.78\pm0.20$	&	$-8.57\pm0.54$							           & $0.16\pm0.06$	\\
$2.69\pm0.08$	&	$1.02\pm0.34$	&	$-9.38\pm0.87$							           & $0.34\pm0.10$ \\
\hline												\hline
\end{tabular}
\end{threeparttable}
\label{tab:no3}
\end{table}

To select the GRB sub-samples at the same $z$ with well constrained best-fit parameters, we impose that the corresponding relations must span at least $2$ order of magnitude in the Combo plane.
Since in each sub-sample the GRB luminosity distances $d_l$ are approximately the same, instead of using $L_0=4\pi d_l^2 F_0$, we employ the rest-frame $0.3$--$10$ keV energy flux $F_0$, rendering this procedure cosmology-independent.
We found seven sub-samples, shown in Fig.~\ref{fig:no2}.
Their best-fit slope parameters $\gamma_z$ (see Table~\ref{tab:no3}) exhibit no evident trend with $z$ within the errors, implying that the Combo relation does not result from an evolution effect. This feature allows us to assume that the above seven sub-samples do follow the same correlation with the same $\gamma$ and $\sigma_{\rm ex}$ but different normalizations. Therefore, we perform a simultaneous fit of the above sub-samples requiring $\gamma$ and $\sigma_{\rm ex}$ to be the same for all of them. The best-fit values, which minimize the sum of the seven log-likelihood functions, are $\gamma=0.90\pm0.13$ and $\sigma_{\rm ex}=0.28\pm0.03$ (see Fig.~\ref{fig:no2}).

\subsubsection{The calibration of $A$}\label{sec:2.2.2}

The Combo relation distance modulus is \citepalias[see, e.g.,][]{Izzo2015},
\begin{align}
\nonumber
\mu_{\rm GRB}=-97.45+\frac{5}{2}&\left[\log\left(\frac{A}{{\rm erg/s}}\right)+\gamma\log \left(\frac{E_{\rm p,i}}{{\rm keV}}\right) -\log\left(\frac{\tau/{\rm s}}{|1+\alpha|}\right)\right. \\
\label{eq:no3}
&\left. -\log\left(\frac{F_0}{\textnormal{erg/cm$^2$/s}}\right)-\log4\pi
\right]\,.
\end{align}
To calibrate $A$ we shall use the nearest GRBs of our sample: 1) 171205A at $z=0.0368$, 2) 180728A at $z=0.117$, and 3) 130702A at $z=0.145$ and 161219B at $z=0.1475$, almost at the same distance.
In Eq.~\ref{eq:no3} we replace $\mu_{\rm GRB}$ with the average distance modulus $\langle\mu_{\rm SNIa}\rangle$ computed from SNe Ia at the same $z$ of the above four sources \citep[][]{Union2,Union2.1}, and we use the value of $\gamma$ from Sec.~\ref{sec:2.2.1} and the values of $F_0$, $E_{\rm p,i}$, $\tau$, and $\alpha$ for each of the four GRBs from Table~\ref{tab:no}. We infer: 1) $\langle\mu_{\rm SNIa}\rangle=35.97\pm0.17$ and $\log[A/{\rm ( erg/s)}]=48.61\pm0.42$ for GRB 171205A; 2) $\langle\mu_{\rm SNIa}\rangle=38.68\pm0.09$ and $\log[A/{\rm ( erg/s)}]=48.97\pm0.25$ for GRB 180728A; 3) $\langle\mu_{\rm SNIa}\rangle=39.21\pm0.24$ and $\log[A/{\rm ( erg/s)}]=49.52\pm0.21$ for GRB 130702A and $\log[A/{\rm ( erg/s)}]=50.09\pm1.12$ for GRB 161219B. Indeed, in estimating the normalization, we select the values of $A$ coming from GRBs 130702A and 161219B because they represent a small set of bursts at the same $z$.\footnote{ Single GRB estimates are disfavored because they might be above/below the relation and, thus, over/underestimate the value of $A$. Small sample estimates in general average these effects.} Their weighted average value is $\log[A/{\rm ( erg/s)}]=49.54\pm0.21$.

\begin{figure}
\centering
\includegraphics[width=\hsize,clip]{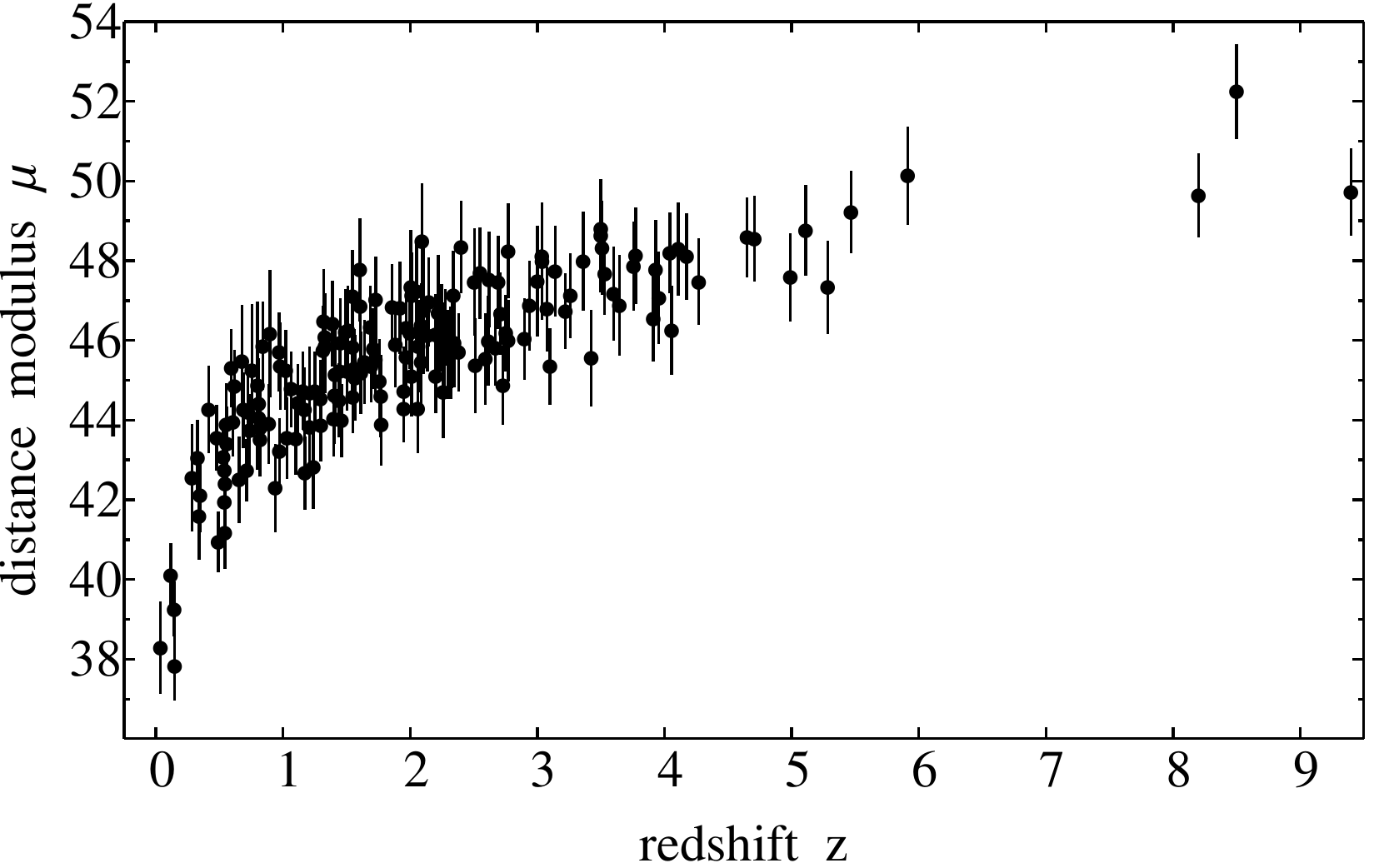}
\caption{The GRB Hubble diagram of the Combo relation.}
\label{fig:no5a}
\end{figure}

\section{Testing the standard $\Lambda$CDM model}\label{sec:3}

GRB distance moduli $\mu_{\rm GRB}$ and their uncertainties $\sigma\mu_{\rm GRB}$ (see Fig.~\ref{fig:no5a}) depend upon the set of observables $(E_{\rm p,i},\tau,\alpha,F_0)$ and therelation parameters. They can be compared with the theoretical distance moduli $\mu_{\rm th}=-97.45+5\log\{d_l[z,\Omega_m,\Omega_\Lambda,w(z),H_0]/{\rm cm}\}$, where the luminosity distance $d_l$ is defined as \citep[see, e.g.,][]{Goobar1995}
\begin{equation}
\label{eq:no5}
d_l = \frac{c}{H_0}\frac{(1+z)}{\sqrt{| \Omega_k |}} F_k\left(\int_0^z \frac{\sqrt{| \Omega_k |} dz^\prime}{\sqrt{E\left[z^\prime,\Omega_m,\Omega_\Lambda, w\right]}}\right)\ ,
\end{equation}
with $E\left[z,\Omega_m,\Omega_\Lambda,w\right]=\Omega_m(1+z)^3+\Omega_\Lambda(1+z)^{3(1+w)}+\Omega_k(1+z)^2$, $\Omega_k = 1-\Omega_m-\Omega_\Lambda$, and $F_k(x)=\sinh(x)$ for $\Omega_k>0$, $F_k(x)=x$ for $\Omega_k = 0$, and $F_k(x)=\sin(x)\}$ for $\Omega_k < 0$.

\begin{table}
\setlength{\tabcolsep}{0.75em}
\renewcommand{\arraystretch}{1.2}
\centering
\caption{Best-fit parameters with $1$--$\sigma$ uncertainties for the various cosmological cases discussed in this work. The last column lists the values of the $\chi^2$. $H_0$ is fixed to the values given by \citetalias{Planck2018} and \citetalias{2019ApJ...876...85R} (see text).}
\small
\begin{tabular}{lccccc}
\hline\hline
$H_0$ &  $\Omega_m$       &  $\Omega_{\Lambda}$  & $\Omega_k$    & $w$   &  $\chi^2$         \\
\hline
\hline
\multicolumn{6}{c}{Flat $\Lambda$CDM (${\rm DOF}=173$)}\\
\cline{1-6}
P20	   &   $0.32^{+0.05}_{-0.05}$   &  $0.68^{+0.05}_{-0.05}$   & $0$   &  $-1$   &  $165.54$  \\
R19   &   $0.22^{+0.04}_{-0.03}$   &  $0.78^{+0.04}_{-0.03}$   & $0$   &  $-1$   &  $171.32$  \\
\hline
\hline
\multicolumn{6}{c}{$\Lambda$CDM (${\rm DOF}=172$)} \\
\cline{1-6}
P20	   &   $0.34^{+0.08}_{-0.07}$   &  $0.91^{+0.22}_{-0.35}$   & $-0.24^{+0.24}_{-0.35}$  &  $-1$    &  $164.38$  \\
R19   &   $0.24^{+0.06}_{-0.05}$   &  $1.01^{+0.15}_{-0.25}$   & $-0.24^{+0.16}_{-0.25}$  &  $-1$   &  $169.23$ \\
\hline
\hline
\end{tabular}
\label{tab:no5}
\end{table}

\begin{figure*}
\centering
\includegraphics[width=0.9\hsize,clip]{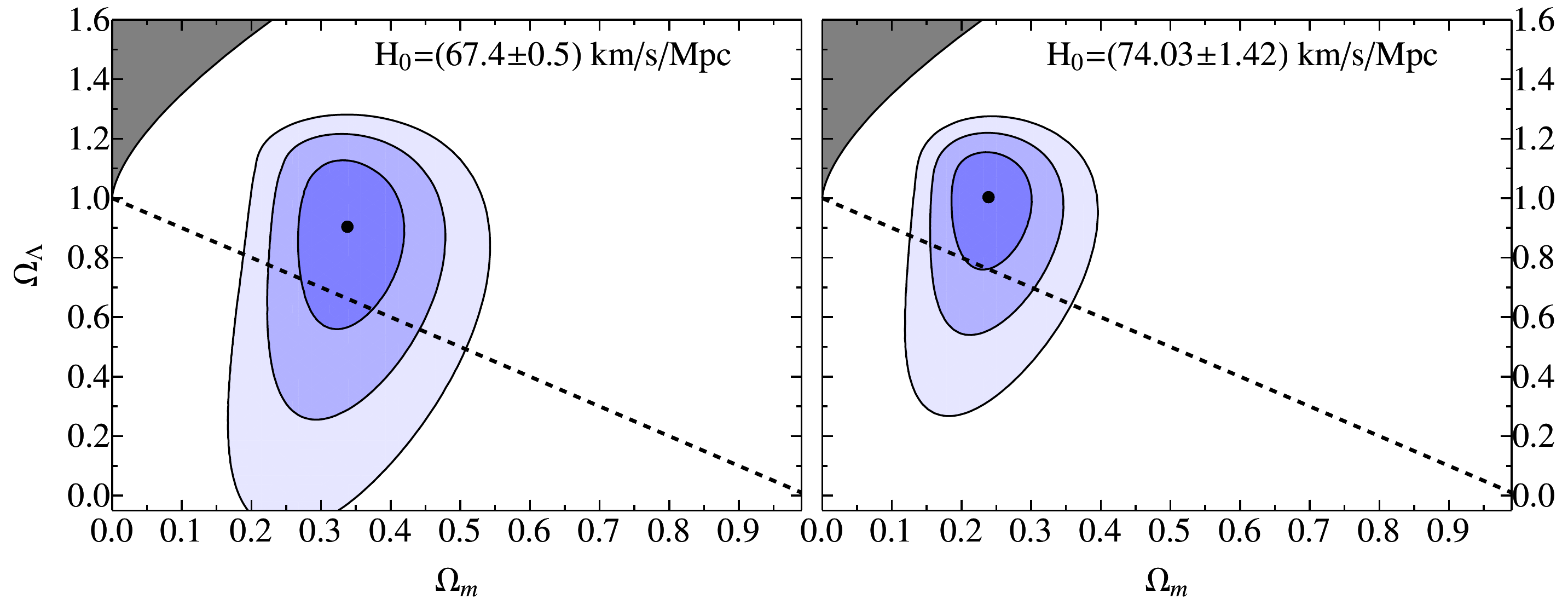}
\caption{Best-fit parameters (black circles) and confidence regions ($1$-, $2$-, and $3$-$\sigma$ from the inner/darker to the outer/lighter) in the $\Omega_m$--$\Omega_\Lambda$ plane for the selected $H_0$. The gray regions indicate the set of parameters for which no Big Bang occurs while the dashed lines those of a flat Universe case.}
\label{fig:no5}
\end{figure*}

We first constrain the Hubble constant $H_0$ with two recent estimates, in tension at $4.4$-$\sigma$: $H_0=(67.4\pm0.5)$~km/s/Mpc \citepalias{Planck2018} and $H_0=(74.03\pm1.42)$~km/s/Mpc \citep[][hereafter R19]{2019ApJ...876...85R}.
Then we compare $\mu_{\rm GRB}$ and $\mu_{\rm th}$ distributions and find the best-fit cosmological parameters by performing a $\chi^2$ statistic test.

We first take the flat $\Lambda$CDM model and then we leave $\Omega_k$ free to vary. In the first case we impose $\Omega_{\Lambda}=1-\Omega_m$ and we summarize our findings in Table~\ref{tab:no5}. In particular,
GRBs provide: $\Omega_m=0.32^{+0.05}_{-0.05}$ for the $H_0$ of \citetalias{Planck2018} and $\Omega_m=0.22^{+0.04}_{-0.03}$ for the $H_0$ of \citetalias{2019ApJ...876...85R}, whereas fits using the non-flat $\Lambda$CDM model provide large values of $\Omega_{\rm k}$ which are consistent within the errors with the flat $\Lambda$CDM case (see Table~\ref{tab:no5} and Fig.~\ref{fig:no5}). Our large and negative value of the curvature parameter is compatible with the recent one obtained by \citet{2020ApJ...888...99W} using only quasar data calibrated via observational Hubble data set (OHD). However, in \citet{2020ApJ...888...99W} the best-fit value $\Omega_{\rm k}=-0.918\pm0.429$ has a large uncertainty, due to the error propagations induced by anchored OHD, and it is in disagreement with the concordance model at $2.1$--$\sigma$ level and at $1.6$--$\sigma$ level with respect to our findings.

Worth noticing, the best-fit $\Omega_m$ values listed in Table~\ref{tab:no5}, though obtained by using different values of $H_0$, are consistent with each other.

\section{Piecewise reconstruction of $w(z)$ in  flat background cosmology}\label{sec:3.4}

We consider now a flat Universe with an evolving DE EoS. We adopt a redshift-binned parametrization for $w(z)$: a step function with different and constant values of $w$ in each of the redshift bins over which the entire redshift range covered by GRBs is divided \citep{Daly2004,King2014}.

The number of bins is arbitrary defined but holds a set of heuristic assumptions that we made before deciding the redshift interval for each bin. First, we have to notice that the number of bins requires the number of points that is enough to guarantee a suitable interpolation. Second, we expect the number of points increases with the redshifts since most of the GRBs of the sample are located at larger redshifts. This implies that we cannot choose tighter intervals as $z$ increases to include always the same number of data points. Rather than following this strategy, we presume to work at least with 15 points for each bins and to split the intervals in redshift differences of about $\Delta z\sim0.5$. In so doing, we assume the following four intervals for $\Delta z_i$: $\Delta z_1=0.55$, $\Delta z_2=0.63$, $\Delta z_3=0.56$ and $\Delta z_4=0.81$, whereas we assume the last interval up to $z\sim 9.5$ since in it we count 48 data points. In this respect the number of data for each interval only slightly increases, leaving unaltered the statistical significance of our fits for each interval. The bins are therefore statistically supported and reported in Table~\ref{tab:no6}.

In Eq.~\ref{eq:no5} we can set
\begin{align}
E\left[\Omega_m,f(z_{N-1}<z \leq z_N)\right]=\,\, & \Omega_m(1+z)^3+\\
&(1-\Omega_m)f(z_{N-1}<z \leq z_N)\,, \nonumber
\end{align}
with parametric  DE evolution $f(z_{N-1}<z \leq z_N)$ given by
\begin{equation}
\label{eq:fzbinned}
f(z_{N-1}<z \le z_N)=(1+z)^{3(1+\tilde{w}_N)}\prod_{i=0}^{N-1}\left[1+{\rm max}(z_i)\right]^{3(\tilde{w}_i-\tilde{w}_{i+1})},
\end{equation}
in which $N$ is the redshift bin where $z$ resides and $\tilde{w}_i$ the EoS parameter in the $i^{\mathrm{th}}$ redshift bin defined by its maximum redshift ${\rm max}(z_i)$.
This parametrization makes no assumptions on DE's nature since different parameters are introduced in every redshift bin.
However, the parameters $w_i$ are usually correlated with each other (see Eq.~\ref{eq:fzbinned}).
This means that the covariance matrix
\begin{equation}
\textbf{C}=\langle \tilde{\textbf{w}} \tilde{\textbf{w}}^{\rm T}\rangle-\langle\tilde{\textbf{w}}\rangle\langle\tilde{\textbf{w}}^{\rm T}\rangle\ ,
\end{equation}
computed from the vector $\tilde{\textbf{w}}$ with components $w_i$ and the transposed one $\tilde{\textbf{w}}^{\rm T}$, is not diagonal.
Following \citet{Huterer2005}, to uncorrelate the DE parameters $\tilde{w}_i$ we transform them through an orthogonal matrix rotation \textbf{O} into a basis that diagonalizes the inverse of the covariance matrix, i.e., ${\bf C}^{-1}={\bf O}^{\rm T}{\bf \Lambda} {\bf O}$, where ${\bf\Lambda}$ is diagonal. The new set of uncorrelated DE parameters $w_i$ is then given by applying a weight matrix defined as $\tilde{\textbf{W}}=\textbf{O}^{\rm T}{\bf \Lambda}^{1/2}\textbf{O}$ to the correlated $\tilde{w}_i$. $\tilde{\textbf{W}}$ is normalized in such a way that its rows, the weights for $\tilde{w}_i$, sum to unity.

\begin{figure*}
\centering
\includegraphics[width=0.9\hsize,clip]{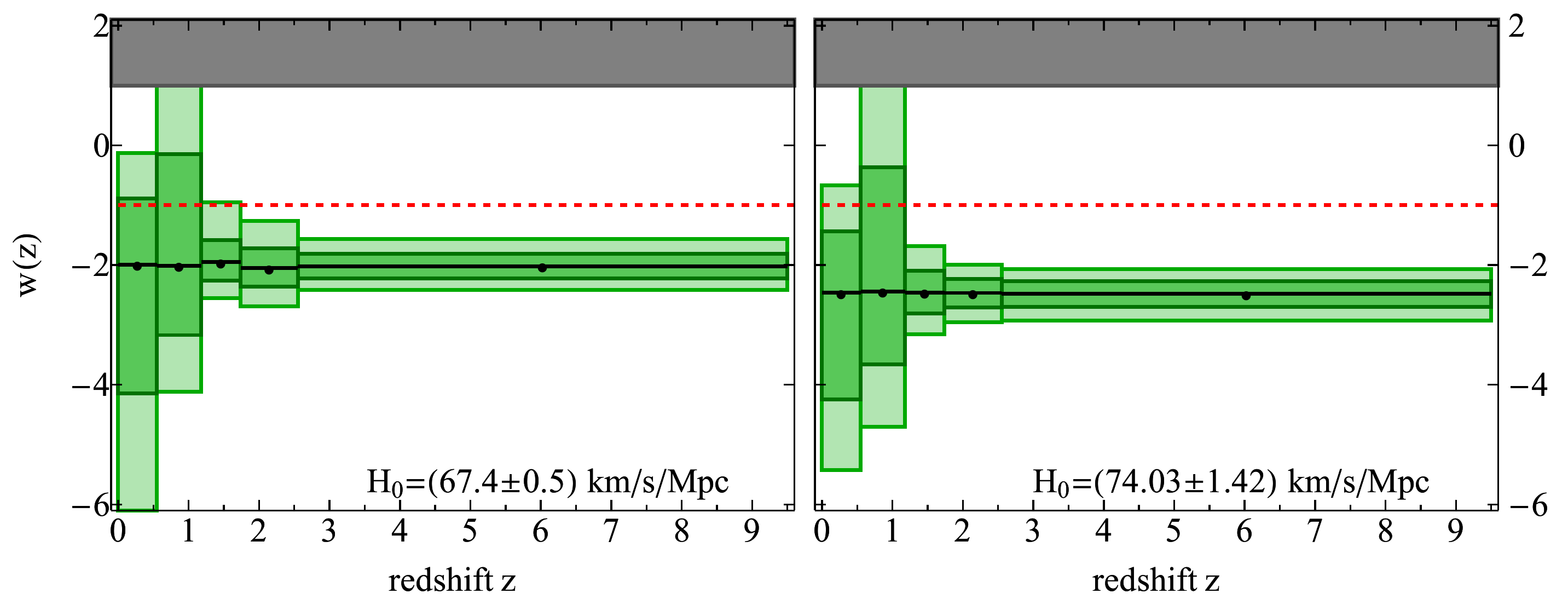}
\caption{The DE EoS reconstructed evolution through the redshift-binned parametrization of $w(z)$ ($1$- and $2$--$\sigma$ from the inner/darker to the outer/lighter) for the selected $H_0$. The dashed red lines mark the value $w=-1$ in the flat $\Lambda$CDM model. The darker region shows unphysical EoS, i.e. exceeding the stiff matter regime.}
\label{fig:no6}
\end{figure*}

We first marginalize over the $\Omega_m$ parameter using the values inferred for the flat $\Lambda$CDM case (see Table~\ref{tab:no5}).
Then, we determine five redshift bins to assure a sufficient number of GRB events for each of them (see Table~\ref{tab:no6}).
The average vector $\langle\tilde{\textbf{w}}\rangle$ is calculated by letting $\tilde{\textbf{w}}$ run over the above marginalization over $\Omega_m$.
The results of the analysis are summarized in Table~\ref{tab:no6} and Fig.~\ref{fig:no6}.
It is worth noting that, for both the values of $H_0$, the $w_i$ are consistent with the value $w=-1$ for $z\lesssim1.2$.
\begin{table}
\renewcommand{\arraystretch}{1.2}
\centering
\caption{Upper part: redshift bin $z_i$, the number of GRBs $N_i$ within, and the corresponding best-fit uncorrelated $w_i$ (with $1$--$\sigma$ attached errors) for the $H_0$ of \citetalias{Planck2018} and \citetalias{2019ApJ...876...85R}. Lower part: values of the $\chi^2$ over the DOF.}
\begin{tabular}{c|c|cc}
\hline\hline
\multirow{2}{*}{$z_i$} &  \multirow{2}{*}{$N_i$}   & \multicolumn{2}{c}{$w_i$}  \\
\cline{3-4}
                      &  &   P20                                  &   R19                    \\
\hline
\hline
$0.00$--$0.55$   & $16$ &   $-1.99^{+1.10}_{-2.15}$   &   $-2.46^{+1.03}_{-1.78}$\\
$0.55$--$1.18$    & $32$&   $-2.01^{+1.86}_{-1.16}$    &   $-2.44^{+2.07}_{-1.22}$\\
$1.18$--$1.74$   & $35$ &   $-1.95^{+0.37}_{-0.31}$    &   $-2.46^{+0.36}_{-0.35}$\\
$1.74$--$2.55$    & $43$&   $-2.05^{+0.33}_{-0.31}$    &   $-2.47^{+0.23}_{-0.24}$\\
$2.55$--$9.50$    & $48$&   $-2.02^{+0.21}_{-0.20}$    &   $-2.48^{+0.21}_{-0.22}$\\
\hline\hline
$\chi^2$/DOF &  &   $168.02/169$   & $161.12/169$\\
\hline
\hline
\end{tabular}
\label{tab:no6}
\end{table}

\section{Theoretical discussions over $w$ evolution}\label{sec:7}

Our aim is to use the updated sample of $174$ GRBs alone in order to get bounds over  DE evolution. Consequently we obtain tighter constraints on the cosmological parameters, compared to previous results in \citetalias{Izzo2015}, where a smaller GRB sample has been used. This first result certifies the goodness of our Combo relation compared with other calibration functions, where the increase of data points does not forcedly lead to improve the quality of fits (Luongo \& Muccino, in publication).

Moreover, we here investigate the idea of fitting cosmological models with GRBs only, i.e. without considering surveys of standard candles. This procedure resembles the analogous method proposed by \citet{2019NatAs...3..272R}, where the authors investigated how to get constraints in cosmology by using only quasars, although a scaling between the Hubble diagram of quasars with that of SNe in the common redshift range $z=0.1$--$1.44$ has been applied. Even though GRBs are not genuine standard candles, their use in cosmology leads to promising outcomes and seems to show a particular behavior constraining $\Omega_m$. In particular, from our findings we first conclude that the higher the $H_0$ value,
\begin{itemize}
    \item[{\bf 1.}] the lower the best-fit value of $\Omega_m$, and
    \item[{\bf 2.}] the better the accuracy on cosmological parameters.
\end{itemize}
However, in all our fits the outcomes have been jeopardized by the common issues related to the use of GRBs in background cosmology \citep{orlando9}. A few of these caveats are discussed below in view of our approach.

\subsection{Issues with low-$z$ GRBs}

The shortage of very low-$z$ GRBs prevents us to obtain a precise value of $H_0$ using GRBs alone. In other words, GRBs alone are unable to fix $H_0$ and to heal somehow the net tension between recent measurements on it, see e.g. \citetalias{Planck2018} and \citetalias{2019ApJ...876...85R}. This is essential for our fits, as if one has a well-bounded $H_0$ the corresponding findings become stabler, implying that $H_0$ plays the role of \emph{an initial value} that fixes the shapes of Hubble curves used in the fit \citep{orlando5,orlando6, orlando7}.

Thus, the lacking of GRBs at small redshift  severely acts on our fits. In fact, at $z\simeq0$ we have no possibility to definitively conclude that the correct cosmological model is the $\Lambda$CDM paradigm.

This means that our approach turns out to be more predictive at intermediate redshifts, i.e. as $z>1.2$. Motivated by this fact, we fixed $H_0$ in our analyses. To do so, we considered the two most recent measurements of $H_0$ by \citetalias{Planck2018} and \citetalias{2019ApJ...876...85R} in tension between each other. It is needful to fix the kinematics of our fit for the above reasons, i.e. we do not have enough GRB sources at small redshifts.

Consequently our findings in the windows of $z<1.2$ permits to take the $\Lambda$CDM suite but leaves open the possibility that DE evolves. Other probes certify slight departures from the case $w=-1$, suggesting that our wide window found using GRBs only at $z=0$ is only due to the lack of sources in this regime.

Going further with the redshift, i.e. as $z>1.2$, leads to refined limits over the DE evolution as we discuss below. Precisely, concerning the redshift-evolving DE EoS, we find that at redshift $z\lesssim1.2$ the evolution of $w$ is almost consistent with the cosmological constant case within $1$--$\sigma$ for both the $H_0$ of \citetalias{2019ApJ...876...85R} and \citetalias{Planck2018}.

\subsection{Dark energy evolution at high $z$}

A first evident deviation from the above-discussed case arises at $z\gtrsim1.5$, where departures from the $\Lambda$CDM case lies on $\gtrsim2.4$--$\sigma$ for the $H_0$ of \citetalias{Planck2018} and on $\gtrsim4$--$\sigma$ for the $H_0$ of \citetalias{2019ApJ...876...85R}. Clearly, at $z\gtrsim1.5$ a severe observational bias exists and we could indeed observe only the brightest GRBs \citep{2020arXiv201002935B} \footnote{Further analysis on this aspect, including the systematics on the entire sample, will be discussed elsewhere.}.

It is evident that the shortage of GRBs at low redshifts has a strong influence for the estimate of the DE EoS parameters $w$. Indeed, DE has its large effects in the low-$z$ Universe ($z \lesssim 0.5$), where $\approx9\%$ of the Combo-GRBs are located. Moreover, while at low-$z$ the DE drives the evolution of the Universe, at $z\geq 1$ the matter dominates. In this light, the epoch of the transition between a DE and a matter-dominated Universes can represent a new constrain for cosmological models (Izzo et al. in preparation).

The Combo-GRB sample and analysis confirm a Universe dominated by  DE at low redshifts, but seems to show a fraction of DE to act even at higher redshifts. Although our data analysis within the flat $\Lambda$CDM model is in agreement with other recent results from different probes (e.g. \citetalias{Planck2018} and \citetalias{2019ApJ...876...85R}), our results prompt a strongly negative mean value for the evolving EoS at intermediate $z$.

In this case, DE may exhibit a phantom behavior, being a likely though troublesome DE candidate. In particular, at larger $z$ current data do not allow to exclude that $w(z)$ might vary with the redshift, possibly to values $w\lesssim-1$. This finding is somehow in line with recent results from \citet{2019NatAs...3..272R}, supporting values of the DE parameter $w<-1$, i.e. a DE density increasing with time, for $z\gtrsim1.4$. Phrasing it differently, we find that the cosmological evolution is driven by DE in lieu of a pure $\Lambda$. By construction the term that dominates is, however, strongly negative, i.e. its evolution shows $w<-1$. As a desirable consequence of this scenario, such a consideration leads to a Universe that begins at Big-Bang and terminates at Big-Rip  \citep{orlando10}.
Indeed, if phantom energy continues driving the universe outward, it could literally tear space into shreds, leading to an unvoidable big rip, as pointed out in \citet{bigrip}.

\begin{figure*}
\centering
\includegraphics[width=0.9\hsize,clip]{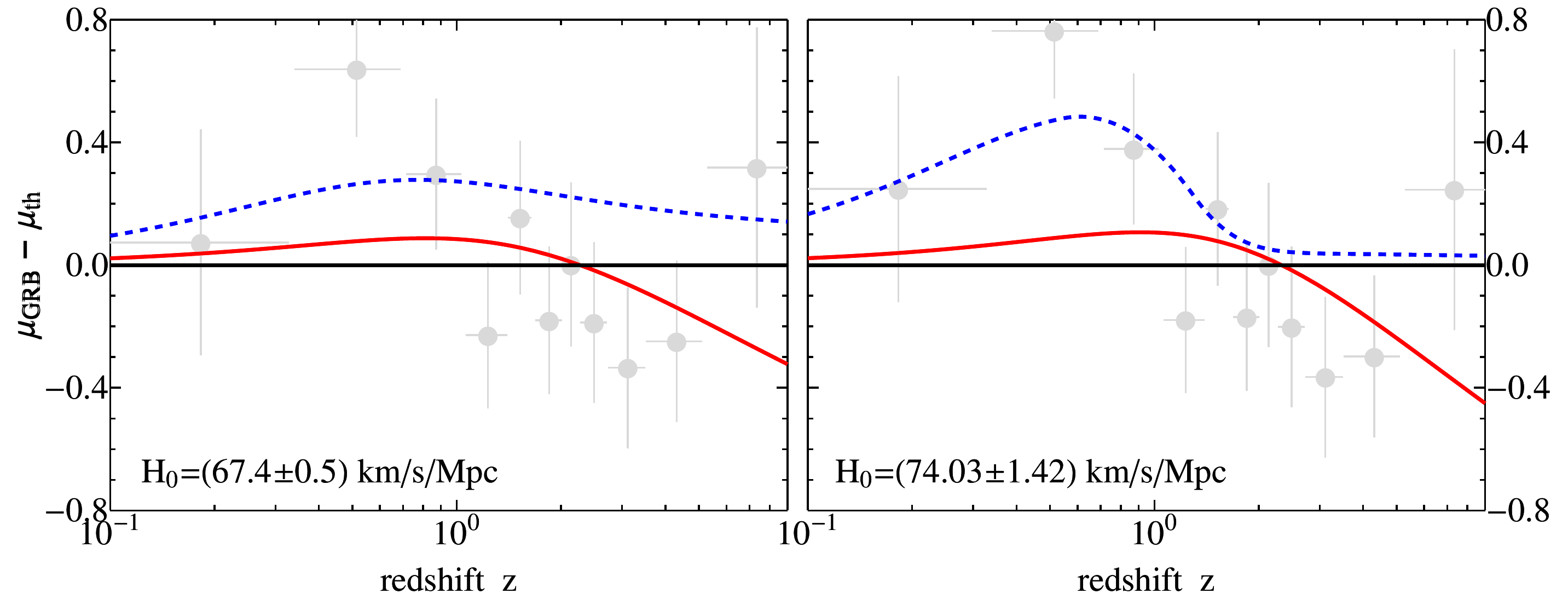}
\caption{The rebinned GRB moduli $\mu_{\rm GRB}$ (gray circles) for the $H_0$ in \citetalias{Planck2018} and \citetalias{2019ApJ...876...85R} compared with the flat $\Lambda$CDM model $\mu_{\rm th}$ (see the horizontal black line and Tab.~\ref{tab:no5}), the non-flat $\Lambda$CDM case (see the red line and Tab.~\ref{tab:no5}), and the flat model with an evolving $w(z)$ (see the dashed blue line and Tab.~\ref{tab:no6}).}
\label{fig:no12}
\end{figure*}

\vspace{1cm}

Although appealing, this scenario is unlikely and turns out to be misleading for essentially two main reasons. First, the $\Lambda$CDM model lies inside our predictions, within the $1$--$\sigma$. Hence, the simplest framework is clearly favored than any deviations. Second, it is arguable that a phantom DE is disfavored and the fact that DE stabilizes its evolution as $z$ increases indicates the need of refined initial values, as above stated. In other words, although the Combo relation improves the quality of overall analyses with GRBs, the lacking of such sources at small redshifts influences the analysis as $z>1.2$.

Summing up the cosmological constant model is still inside the GRBs predictions got from the use of the Combo relation. Evident deviations to phantom DE and/or to slight variation from $w=-1$ may be addressed noticing that at $z\simeq0$ no enough GRB sources are available.

\subsection{Does the cosmic speed up behave like phantom?}

The overall fitting procedures that we considered are clearly dependent on the choice of the data sets and bins here used. In particular, using GRBs alone likely leads to $w<-1$, while adopting combined data sets would significantly help to refine the numerical outcomes.
However, involving more than one data set based on GRBs, although it provides consistency with \citetalias{Planck2018},  the errors are suitably large, with significant discrepancy. The case is similar to what found for quasars, since treated separately, there is no tension, but as one attempts to anchor them using another distance indicator \citep{2020ApJ...888...99W}, one gets issues on errors. Thus, the indication $w<-1$ provided by GRBs can be refined but with further error propagations induced by anchored further data surveys. A relevant approach to overcome this issue is to enlarge the whole data set adopting alternative strategies of experimental analysis \citep[see, e.g.,][]{LM2020}.

\section{Final remarks}\label{sec:8}

In this work, we considered an extended GRB sample, characterized by 174 GRBs and we fixed limits over DE's evolution using the Combo relation. We extended the approach made in \citetalias{Izzo2015} and we got refined bounds over the whole matter content, i.e. $\Omega_m$ and DE density. We first worked with the $\Lambda$CDM model and then we constrained the evolution of $w(z)$ through a piecewise formulation over seven GRB redshift interval bins. We theoretically discussed our results, showing an overall agreement with the $\Lambda$CDM model that lies inside our findings. We distinguished two regimes, $z<1.2$ and $z\geq1.2$, where GRBs turn out to be more predictive. Deviations from $w=-1$ are not excluded in both the intervals. However, we proposed that at small redshifts, deviations from $w=-1$ are jeopardized by the GRB shortage. As the redshift increases, DE's abundance is much more constrained with GRBs. Thus, beyond $z\simeq1.2$ the case $w<-1$ cannot be fully excluded from our computations.

In particular, a concise summary of the fitting of $\mu_{\rm GRB}$ with various cosmological models has been  shown in Fig.~\ref{fig:no12}, where we compared the residual GRB distance moduli $\mu_{\rm GRB}$ with respect the $\mu_{\rm th}$ from the flat $\Lambda$CDM case (see horizontal black line and Tab.~\ref{tab:no5}). In Fig.~\ref{fig:no12} we also compared these figures with the distance moduli from the non-flat $\Lambda$CDM case (see the red line and Tab.~\ref{tab:no5}), and the flat wCDM with an evolving $w(z)$ (see the dashed blue line and Tab.~\ref{tab:no6}).

From all our analyses we obtained the following main results:
\begin{itemize}
\item[1)] the Combo-GRB sample confirmed a DE-dominated Universe, responsible for the observed acceleration at low redshifts. Our data analysis within the flat $\Lambda$CDM model is in agreement with other recent results from different probes (e.g. \citetalias{Planck2018} and \citetalias{2019ApJ...876...85R});
\item[2)] as stated above, it was clear that the shortage of GRBs at low redshifts has a strong influence for the estimate of  $w$. This happens since  DE has its large effects around $z \lesssim 0.5$, where $\approx9\%$ of the Combo-GRBs are located;
\item[3)] the value of w$(z)$ agreed, within 1$-\sigma$, with the standard $\Lambda$CDM model up to $z\approx1.2$. At larger $z$, current data do not allow to exclude that $w(z)$ might vary with the redshift, possibly to values $w\lesssim-1$. This finding is somehow in line with recent results from \citet{2019NatAs...3..272R}, supporting values of the DE parameter $w<-1$, i.e. a DE density increasing with time, for $z\gtrsim1.4$.
\end{itemize}

From the above considerations, we concluded by remarking the important role of GRBs as probes for the low- and high-$z$ Universe: the evidence of tight correlations between their observed properties has really become crucial to understand the evolution history of the Universe at very large redshifts, where no astrophysical probe has gone before.

Future perspectives are based on missions like the incoming SVOM and the proposed eXTP and THESEUS that will provide a robust dataset of GRBs at all redshifts ($\approx100$ at $z\gtrsim5$ over a three year mission) allowing a more detailed study of the early Universe up the re-ionization era.
The application of this forthcoming dataset to further test the Combo relation can also shed light on the physics of the X-ray afterglow and provide a physical explanation behind this GRB relation \citep[see, e.g.,][]{2017MNRAS.468..570M,2018arXiv180408652S}.


\begin{acknowledgements}
The authors are thankful to the anonymous referee for constructive suggestions that helped to significantly improve the manuscript.
This work made use of data supplied by the UK \emph{Swift} Science Data Center at the University of Leicester.
The work was partially supported by the Ministry of Education and Science of the Republic of Kazakhstan,  Grant: IRN AP08052311.
MM is supported by INFN as part of the MoonLIGHT-2 experiment in the framework of the research activities of CSN2.
LA acknowledges financial contribution from the agreement ASI-INAF n.~2017-14-H.O.
We warmly thank Michael Oliver for the support in completing this paper. 
\end{acknowledgements}


\begin{appendix}

\section{The sample}\label{appendix1}

In Table~\ref{tab:no} we list the full sample long GRBs used in this work and their Combo relation parameters.
%
\footnotesize
\LTcapwidth=\linewidth
\begin{longtable}{lcccccc}
\caption{List of the long bursts analyzed in \citetalias{Izzo2015} and in this work (in boldface) and their Combo relation parameters. GRBs belonging to the sub-samples at $\langle z\rangle\approx0.54$, $1.18$, $1.46$, $1.70$, $2.05$, $2.27$, and $2.69$ are labeled by $\textit{a}$, $\textit{b}$, $\textit{c}$, $\textit{d}$, $\textit{e}$, $\textit{f}$, and $\textit{g}$, respectively. Errors are at 1 $\sigma$ confidence level.} \label{tab:no}\\
\hline\hline
GRB      &  z        &  Sub-sample  &  $\log \left(E_{\rm p,i}/{\rm keV}\right)$  &  $\log \left(F_0/{\rm erg}/{\rm cm}^2{\rm s}\right)$  &  $\log \left(\tau/{\rm s}\right)$     &  $\alpha$ \\
\hline
\endfirsthead
\caption{continued.}\\
\hline\hline
GRB      &  z        &  Sub-sample  &  $\log \left(E_{\rm p,i}/{\rm keV}\right)$  &  $\log \left(F_0/{\rm erg}/{\rm cm}^2{\rm s}\right)$  &  $\log \left(\tau/{\rm s}\right)$     &  $\alpha$ \\
\hline
\endhead
\hline
\endfoot
	050401		&	$	2.8983	$	&		&	$	2.67	^	{	+	0.13	}	_	{	-0.11	}	$	&	$	-9.57	^	{	+	0.02	}	_	{	-0.02	}	$	&	$	2.90	^	{	+	0.06	}	_	{	-0.07	}	$	&	$	-1.75	\pm	0.09	$	\\
	050922C		&	$	2.1995	$	&	\textit{f}	&	$	2.62	^	{	+	0.13	}	_	{	-0.10	}	$	&	$	-8.79	^	{	+	0.05	}	_	{	-0.06	}	$	&	$	1.64	^	{	+	0.07	}	_	{	-0.09	}	$	&	$	-1.30	\pm	0.03	$	\\
	051109A		&	$	2.346	$	&		&	$	2.73	^	{	+	0.13	}	_	{	-0.19	}	$	&	$	-9.86	^	{	+	0.07	}	_	{	-0.09	}	$	&	$	2.97	^	{	+	0.08	}	_	{	-0.10	}	$	&	$	-1.36	\pm	0.03	$	\\
	060115		&	$	3.53	$	&		&	$	2.47	^	{	+	0.05	}	_	{	-0.05	}	$	&	$	-11.15	^	{	+	0.06	}	_	{	-0.06	}	$	&	$	3.99	^	{	+	0.13	}	_	{	-0.19	}	$	&	$	-2.64	\pm	0.44	$	\\
	060418		&	$	1.489	$	&	\textit{c}	&	$	2.76	^	{	+	0.10	}	_	{	-0.12	}	$	&	$	-9.22	^	{	+	0.07	}	_	{	-0.08	}	$	&	$	2.49	^	{	+	0.08	}	_	{	-0.10	}	$	&	$	-1.64	\pm	0.06	$	\\
	060526		&	$	3.2213	$	&		&	$	2.02	^	{	+	0.08	}	_	{	-0.10	}	$	&	$	-11.31	^	{	+	0.04	}	_	{	-0.04	}	$	&	$	4.53	^	{	+	0.14	}	_	{	-0.21	}	$	&	$	-5.23	\pm	1.40	$	\\
	060707		&	$	3.424	$	&		&	$	2.44	^	{	+	0.13	}	_	{	-0.08	}	$	&	$	-10.60	^	{	+	0.14	}	_	{	-0.21	}	$	&	$	2.87	^	{	+	0.19	}	_	{	-0.36	}	$	&	$	-1.07	\pm	0.05	$	\\
	060729		&	$	0.54	$	&	\textit{a}	&	$	1.89	^	{	+	0.12	}	_	{	-0.16	}	$	&	$	-10.54	^	{	+	0.01	}	_	{	-0.01	}	$	&	$	4.80	^	{	+	0.02	}	_	{	-0.02	}	$	&	$	-1.76	\pm	0.03	$	\\
	060814		&	$	0.84	$	&		&	$	2.88	^	{	+	0.12	}	_	{	-0.17	}	$	&	$	-10.42	^	{	+	0.04	}	_	{	-0.04	}	$	&	$	3.88	^	{	+	0.06	}	_	{	-0.07	}	$	&	$	-1.56	\pm	0.07	$	\\
	060908		&	$	1.8836	$	&	\textit{d}	&	$	2.74	^	{	+	0.07	}	_	{	-0.09	}	$	&	$	-8.50	^	{	+	0.07	}	_	{	-0.08	}	$	&	$	1.65	^	{	+	0.10	}	_	{	-0.13	}	$	&	$	-1.37	\pm	0.05	$	\\
	060927		&	$	5.467	$	&		&	$	2.44	^	{	+	0.10	}	_	{	-0.14	}	$	&	$	-10.80	^	{	+	0.06	}	_	{	-0.06	}	$	&	$	2.51	^	{	+	0.11	}	_	{	-0.15	}	$	&	$	-1.55	\pm	0.17	$	\\
	061121		&	$	1.3145	$	&		&	$	3.11	^	{	+	0.05	}	_	{	-0.05	}	$	&	$	-9.52	^	{	+	0.01	}	_	{	-0.01	}	$	&	$	3.23	^	{	+	0.02	}	_	{	-0.02	}	$	&	$	-1.55	\pm	0.02	$	\\
	071020		&	$	2.1462	$	&		&	$	3.01	^	{	+	0.06	}	_	{	-0.07	}	$	&	$	-8.50	^	{	+	0.06	}	_	{	-0.07	}	$	&	$	1.22	^	{	+	0.07	}	_	{	-0.09	}	$	&	$	-1.21	\pm	0.02	$	\\
	080319B		&	$	0.9382	$	&		&	$	3.10	^	{	+	0.01	}	_	{	-0.01	}	$	&	$	-6.13	^	{	+	0.03	}	_	{	-0.03	}	$	&	$	1.29	^	{	+	0.03	}	_	{	-0.03	}	$	&	$	-1.67	\pm	0.01	$	\\
	080413B		&	$	1.1014	$	&	\textit{b}	&	$	2.21	^	{	+	0.05	}	_	{	-0.06	}	$	&	$	-9.15	^	{	+	0.04	}	_	{	-0.04	}	$	&	$	1.97	^	{	+	0.06	}	_	{	-0.07	}	$	&	$	-1.06	\pm	0.02	$	\\
	080605		&	$	1.6403	$	&	\textit{d}	&	$	2.88	^	{	+	0.04	}	_	{	-0.05	}	$	&	$	-8.45	^	{	+	0.02	}	_	{	-0.02	}	$	&	$	1.94	^	{	+	0.03	}	_	{	-0.03	}	$	&	$	-1.41	\pm	0.02	$	\\
	080607		&	$	3.0368	$	&		&	$	3.23	^	{	+	0.05	}	_	{	-0.04	}	$	&	$	-9.72	^	{	+	0.08	}	_	{	-0.09	}	$	&	$	2.73	^	{	+	0.08	}	_	{	-0.10	}	$	&	$	-1.76	\pm	0.07	$	\\
	080721		&	$	2.5914	$	&		&	$	3.24	^	{	+	0.06	}	_	{	-0.06	}	$	&	$	-8.11	^	{	+	0.01	}	_	{	-0.01	}	$	&	$	1.97	^	{	+	0.01	}	_	{	-0.01	}	$	&	$	-1.48	\pm	0.01	$	\\
	080810		&	$	3.3604	$	&		&	$	3.17	^	{	+	0.05	}	_	{	-0.06	}	$	&	$	-9.71	^	{	+	0.13	}	_	{	-0.18	}	$	&	$	2.86	^	{	+	0.13	}	_	{	-0.19	}	$	&	$	-2.07	\pm	0.15	$	\\
	080916A		&	$	0.689	$	&		&	$	2.32	^	{	+	0.04	}	_	{	-0.05	}	$	&	$	-10.32	^	{	+	0.06	}	_	{	-0.07	}	$	&	$	3.09	^	{	+	0.10	}	_	{	-0.12	}	$	&	$	-1.08	\pm	0.04	$	\\
	080928		&	$	1.692	$	&	\textit{d}	&	$	1.98	^	{	+	0.09	}	_	{	-0.12	}	$	&	$	-10.27	^	{	+	0.06	}	_	{	-0.07	}	$	&	$	3.60	^	{	+	0.09	}	_	{	-0.11	}	$	&	$	-2.66	\pm	0.21	$	\\
	081007		&	$	0.5295	$	&	\textit{a}	&	$	1.79	^	{	+	0.10	}	_	{	-0.12	}	$	&	$	-10.19	^	{	+	0.05	}	_	{	-0.05	}	$	&	$	3.20	^	{	+	0.07	}	_	{	-0.09	}	$	&	$	-1.15	\pm	0.04	$	\\
	081008		&	$	1.9685	$	&	\textit{d}	&	$	2.42	^	{	+	0.08	}	_	{	-0.10	}	$	&	$	-10.15	^	{	+	0.04	}	_	{	-0.04	}	$	&	$	3.23	^	{	+	0.07	}	_	{	-0.08	}	$	&	$	-1.93	\pm	0.11	$	\\
	081222		&	$	2.77	$	&	\textit{g}	&	$	2.80	^	{	+	0.04	}	_	{	-0.05	}	$	&	$	-8.52	^	{	+	0.03	}	_	{	-0.03	}	$	&	$	1.45	^	{	+	0.04	}	_	{	-0.04	}	$	&	$	-1.22	\pm	0.01	$	\\
	090102		&	$	1.547	$	&		&	$	3.07	^	{	+	0.04	}	_	{	-0.05	}	$	&	$	-8.38	^	{	+	0.09	}	_	{	-0.11	}	$	&	$	1.93	^	{	+	0.09	}	_	{	-0.11	}	$	&	$	-1.45	\pm	0.03	$	\\
	090418A		&	$	1.608	$	&		&	$	3.20	^	{	+	0.10	}	_	{	-0.12	}	$	&	$	-9.34	^	{	+	0.02	}	_	{	-0.02	}	$	&	$	2.78	^	{	+	0.04	}	_	{	-0.05	}	$	&	$	-1.68	\pm	0.05	$	\\
	090423		&	$	8.2	$	&		&	$	2.65	^	{	+	0.05	}	_	{	-0.05	}	$	&	$	-11.25	^	{	+	0.06	}	_	{	-0.06	}	$	&	$	2.68	^	{	+	0.12	}	_	{	-0.17	}	$	&	$	-1.28	\pm	0.14	$	\\
	090424		&	$	0.544	$	&	\textit{a}	&	$	2.40	^	{	+	0.04	}	_	{	-0.05	}	$	&	$	-8.39	^	{	+	0.03	}	_	{	-0.03	}	$	&	$	2.39	^	{	+	0.03	}	_	{	-0.04	}	$	&	$	-1.21	\pm	0.01	$	\\
	090516		&	$	4.109	$	&		&	$	2.99	^	{	+	0.18	}	_	{	-0.11	}	$	&	$	-10.49	^	{	+	0.05	}	_	{	-0.06	}	$	&	$	3.40	^	{	+	0.08	}	_	{	-0.09	}	$	&	$	-2.20	\pm	0.14	$	\\
	090618		&	$	0.54	$	&	\textit{a}	&	$	2.40	^	{	+	0.04	}	_	{	-0.05	}	$	&	$	-8.89	^	{	+	0.01	}	_	{	-0.01	}	$	&	$	3.06	^	{	+	0.02	}	_	{	-0.02	}	$	&	$	-1.44	\pm	0.01	$	\\
	091018		&	$	0.971	$	&		&	$	1.74	^	{	+	0.13	}	_	{	-0.20	}	$	&	$	-9.03	^	{	+	0.02	}	_	{	-0.03	}	$	&	$	2.24	^	{	+	0.04	}	_	{	-0.05	}	$	&	$	-1.29	\pm	0.02	$	\\
	091020		&	$	1.71	$	&	\textit{d}	&	$	2.71	^	{	+	0.05	}	_	{	-0.06	}	$	&	$	-9.29	^	{	+	0.03	}	_	{	-0.03	}	$	&	$	2.40	^	{	+	0.04	}	_	{	-0.05	}	$	&	$	-1.34	\pm	0.03	$	\\
	091029		&	$	2.752	$	&	\textit{g}	&	$	2.36	^	{	+	0.11	}	_	{	-0.15	}	$	&	$	-10.85	^	{	+	0.02	}	_	{	-0.03	}	$	&	$	3.57	^	{	+	0.05	}	_	{	-0.06	}	$	&	$	-1.39	\pm	0.05	$	\\
	091127		&	$	0.49	$	&		&	$	1.71	^	{	+	0.04	}	_	{	-0.05	}	$	&	$	-8.99	^	{	+	0.03	}	_	{	-0.03	}	$	&	$	3.33	^	{	+	0.03	}	_	{	-0.04	}	$	&	$	-1.52	\pm	0.02	$	\\
	100621A		&	$	0.542	$	&	\textit{a}	&	$	2.16	^	{	+	0.08	}	_	{	-0.06	}	$	&	$	-9.77	^	{	+	0.05	}	_	{	-0.06	}	$	&	$	3.62	^	{	+	0.08	}	_	{	-0.10	}	$	&	$	-1.08	\pm	0.04	$	\\
	100814A		&	$	1.44	$	&	\textit{c}	&	$	2.41	^	{	+	0.05	}	_	{	-0.06	}	$	&	$	-10.76	^	{	+	0.01	}	_	{	-0.02	}	$	&	$	4.85	^	{	+	0.05	}	_	{	-0.05	}	$	&	$	-2.77	\pm	0.16	$	\\
	100906A		&	$	1.727	$	&	\textit{d}	&	$	2.46	^	{	+	0.06	}	_	{	-0.08	}	$	&	$	-9.83	^	{	+	0.04	}	_	{	-0.04	}	$	&	$	3.40	^	{	+	0.05	}	_	{	-0.06	}	$	&	$	-2.27	\pm	0.10	$	\\
	110213A		&	$	1.46	$	&	\textit{c}	&	$	2.35	^	{	+	0.04	}	_	{	-0.04	}	$	&	$	-9.23	^	{	+	0.02	}	_	{	-0.02	}	$	&	$	3.39	^	{	+	0.04	}	_	{	-0.04	}	$	&	$	-2.48	\pm	0.09	$	\\
	110422A		&	$	1.77	$	&		&	$	2.62	^	{	+	0.01	}	_	{	-0.01	}	$	&	$	-8.63	^	{	+	0.08	}	_	{	-0.10	}	$	&	$	2.13	^	{	+	0.09	}	_	{	-0.12	}	$	&	$	-1.33	\pm	0.03	$	\\
	111228A		&	$	0.714	$	&		&	$	1.76	^	{	+	0.05	}	_	{	-0.06	}	$	&	$	-10.10	^	{	+	0.03	}	_	{	-0.03	}	$	&	$	3.68	^	{	+	0.05	}	_	{	-0.06	}	$	&	$	-1.42	\pm	0.04	$	\\
	120119A		&	$	1.728	$	&	\textit{d}	&	$	2.70	^	{	+	0.04	}	_	{	-0.05	}	$	&	$	-10.61	^	{	+	0.08	}	_	{	-0.10	}	$	&	$	3.69	^	{	+	0.13	}	_	{	-0.19	}	$	&	$	-1.99	\pm	0.23	$	\\
	120811C		&	$	2.671	$	&	\textit{g}	&	$	2.30	^	{	+	0.02	}	_	{	-0.02	}	$	&	$	-10.16	^	{	+	0.04	}	_	{	-0.04	}	$	&	$	3.12	^	{	+	0.12	}	_	{	-0.16	}	$	&	$	-1.57	\pm	0.18	$	\\
	120907A		&	$	0.97	$	&		&	$	2.48	^	{	+	0.08	}	_	{	-0.10	}	$	&	$	-10.11	^	{	+	0.04	}	_	{	-0.05	}	$	&	$	2.75	^	{	+	0.07	}	_	{	-0.09	}	$	&	$	-1.17	\pm	0.04	$	\\
	120922A		&	$	3.1	$	&		&	$	2.19	^	{	+	0.04	}	_	{	-0.04	}	$	&	$	-9.81	^	{	+	0.09	}	_	{	-0.11	}	$	&	$	2.26	^	{	+	0.11	}	_	{	-0.15	}	$	&	$	-1.14	\pm	0.04	$	\\
	121128A		&	$	2.2	$	&	\textit{f}	&	$	2.39	^	{	+	0.02	}	_	{	-0.02	}	$	&	$	-9.16	^	{	+	0.03	}	_	{	-0.04	}	$	&	$	2.84	^	{	+	0.07	}	_	{	-0.08	}	$	&	$	-2.27	\pm	0.14	$	\\
	121211A		&	$	1.023	$	&		&	$	2.29	^	{	+	0.05	}	_	{	-0.06	}	$	&	$	-10.57	^	{	+	0.08	}	_	{	-0.09	}	$	&	$	3.41	^	{	+	0.14	}	_	{	-0.20	}	$	&	$	-1.27	\pm	0.11	$	\\
	130408A		&	$	3.757	$	&		&	$	3.00	^	{	+	0.06	}	_	{	-0.07	}	$	&	$	-9.96	^	{	+	0.05	}	_	{	-0.05	}	$	&	$	2.81	^	{	+	0.08	}	_	{	-0.10	}	$	&	$	-1.66	\pm	0.12	$	\\
	130420A		&	$	1.297	$	&		&	$	2.11	^	{	+	0.02	}	_	{	-0.02	}	$	&	$	-10.25	^	{	+	0.06	}	_	{	-0.07	}	$	&	$	2.76	^	{	+	0.10	}	_	{	-0.13	}	$	&	$	-1.05	\pm	0.03	$	\\
	130427A		&	$	0.3399	$	&		&	$	3.06	^	{	+	0.01	}	_	{	-0.01	}	$	&	$	-7.63	^	{	+	0.02	}	_	{	-0.02	}	$	&	$	2.79	^	{	+	0.03	}	_	{	-0.03	}	$	&	$	-1.37	\pm	0.01	$	\\
	130505A		&	$	2.27	$	&	\textit{f}	&	$	3.31	^	{	+	0.02	}	_	{	-0.02	}	$	&	$	-9.30	^	{	+	0.04	}	_	{	-0.04	}	$	&	$	3.24	^	{	+	0.04	}	_	{	-0.05	}	$	&	$	-1.65	\pm	0.03	$	\\
	130610A		&	$	2.092	$	&	\textit{e}	&	$	2.96	^	{	+	0.06	}	_	{	-0.07	}	$	&	$	-9.71	^	{	+	0.24	}	_	{	-0.56	}	$	&	$	1.81	^	{	+	0.20	}	_	{	-0.37	}	$	&	$	-1.23	\pm	0.12	$	\\
	130612A		&	$	2.006	$	&	\textit{e}	&	$	2.27	^	{	+	0.07	}	_	{	-0.08	}	$	&	$	-10.89	^	{	+	0.25	}	_	{	-0.65	}	$	&	$	2.80	^	{	+	0.22	}	_	{	-0.48	}	$	&	$	-1.22	\pm	0.42	$	\\
	130701A		&	$	1.155	$	&	\textit{b}	&	$	2.28	^	{	+	0.02	}	_	{	-0.02	}	$	&	$	-8.60	^	{	+	0.10	}	_	{	-0.12	}	$	&	$	1.65	^	{	+	0.12	}	_	{	-0.16	}	$	&	$	-1.26	\pm	0.03	$	\\
	130702A		&	$	0.145	$	&		&	$	1.17	^	{	+	0.06	}	_	{	-0.07	}	$	&	$	-10.44	^	{	+	0.05	}	_	{	-0.06	}	$	&	$	4.82	^	{	+	0.06	}	_	{	-0.08	}	$	&	$	-1.37	\pm	0.03	$	\\
	130831A		&	$	0.4791	$	&		&	$	1.91	^	{	+	0.03	}	_	{	-0.03	}	$	&	$	-9.62	^	{	+	0.05	}	_	{	-0.06	}	$	&	$	3.25	^	{	+	0.06	}	_	{	-0.07	}	$	&	$	-1.74	\pm	0.07	$	\\
	131030A		&	$	1.295	$	&		&	$	2.61	^	{	+	0.02	}	_	{	-0.02	}	$	&	$	-9.12	^	{	+	0.03	}	_	{	-0.03	}	$	&	$	2.60	^	{	+	0.05	}	_	{	-0.06	}	$	&	$	-1.30	\pm	0.02	$	\\
	131105A		&	$	1.686	$	&	\textit{d}	&	$	2.62	^	{	+	0.09	}	_	{	-0.12	}	$	&	$	-10.42	^	{	+	0.03	}	_	{	-0.04	}	$	&	$	3.15	^	{	+	0.08	}	_	{	-0.10	}	$	&	$	-1.28	\pm	0.08	$	\\
	131117A		&	$	4.042	$	&		&	$	2.35	^	{	+	0.07	}	_	{	-0.08	}	$	&	$	-10.61	^	{	+	0.08	}	_	{	-0.10	}	$	&	$	2.39	^	{	+	0.13	}	_	{	-0.19	}	$	&	$	-1.30	\pm	0.09	$	\\
	140206		&	$	2.73	$	&	\textit{g}	&	$	2.65	^	{	+	0.02	}	_	{	-0.02	}	$	&	$	-9.37	^	{	+	0.02	}	_	{	-0.02	}	$	&	$	2.91	^	{	+	0.03	}	_	{	-0.03	}	$	&	$	-1.44	\pm	0.02	$	\\
	140213		&	$	1.2076	$	&	\textit{b}	&	$	2.25	^	{	+	0.01	}	_	{	-0.01	}	$	&	$	-9.76	^	{	+	0.04	}	_	{	-0.05	}	$	&	$	3.40	^	{	+	0.06	}	_	{	-0.07	}	$	&	$	-1.48	\pm	0.05	$	\\
\textbf{050215B	}	&	$	2.62	$	&	\textit{g}	&	$	1.80	^	{	+	0.13	}	_	{	-0.07	}	$	&	$	-11.79	^	{	+	0.15	}	_	{	-0.24	}	$	&	$	3.47	^	{	+	0.23	}	_	{	-0.52	}	$	&	$	-1.40	\pm	0.28	$	\\
\textbf{050315	}	&	$	1.95	$	&		&	$	1.95	^	{	+	0.08	}	_	{	-0.10	}	$	&	$	-11.05	^	{	+	0.02	}	_	{	-0.03	}	$	&	$	4.45	^	{	+	0.08	}	_	{	-0.10	}	$	&	$	-1.78	\pm	0.17	$	\\
\textbf{050318	}	&	$	1.4436	$	&	\textit{c}	&	$	2.06	^	{	+	0.09	}	_	{	-0.11	}	$	&	$	-10.05	^	{	+	0.09	}	_	{	-0.11	}	$	&	$	3.31	^	{	+	0.12	}	_	{	-0.16	}	$	&	$	-2.07	\pm	0.18	$	\\
\textbf{050505}&	$4.27$		&	&	$2.79^{+0.13}_{-0.18}$	&	$-10.53^{+0.03}_{-0.03}$	&	$3.52^{+0.06}_{-0.07}$	&	$-1.99\pm0.10$ \\
\textbf{050525A	}	&	$	0.606	$	&		&	$	2.11	^	{	+	0.04	}	_	{	-0.05	}	$	&	$	-9.26	^	{	+	0.05	}	_	{	-0.06	}	$	&	$	2.78	^	{	+	0.06	}	_	{	-0.07	}	$	&	$	-1.54	\pm	0.05	$	\\
\textbf{050819A}&	$2.5043$	&	&	$1.89^{+0.30}_{-0.64}$	&	$-11.94^{+0.14}_{-0.21}$	&	$3.62^{+0.26}_{-0.71}$	&	$-1.32\pm0.29$ \\
\textbf{050820A	}	&	$	2.6147	$	&	\textit{g}	&	$	3.12	^	{	+	0.08	}	_	{	-0.10	}	$	&	$	-9.71	^	{	+	0.02	}	_	{	-0.02	}	$	&	$	3.05	^	{	+	0.03	}	_	{	-0.03	}	$	&	$	-1.28	\pm	0.02	$	\\
\textbf{051008	}	&	$	2.77	$	&	\textit{g}	&	$	3.06	^	{	+	0.13	}	_	{	-0.18	}	$	&	$	-10.60	^	{	+	0.03	}	_	{	-0.03	}	$	&	$	4.22	^	{	+	0.12	}	_	{	-0.16	}	$	&	$	-6.02	\pm	1.32	$	\\
\textbf{051022	}	&	$	0.809	$	&		&	$	2.88	^	{	+	0.13	}	_	{	-0.18	}	$	&	$	-9.70	^	{	+	0.07	}	_	{	-0.08	}	$	&	$	4.12	^	{	+	0.07	}	_	{	-0.09	}	$	&	$	-2.35	\pm	0.11	$	\\
\textbf{060111A}&	$2.32$		&	\textit{f}  &	$2.39^{+0.10}_{-0.06}$	&	$-11.27^{+0.12}_{-0.17}$	&	$3.22^{+0.19}_{-0.33}$	&	$-1.04\pm0.07$ \\
\textbf{060124	}	&	$	2.296	$	&	\textit{f}	&	$	2.89	^	{	+	0.13	}	_	{	-0.20	}	$	&	$	-9.99	^	{	+	0.03	}	_	{	-0.04	}	$	&	$	3.49	^	{	+	0.05	}	_	{	-0.05	}	$	&	$	-1.49	\pm	0.04	$	\\
\textbf{060204B}&	$2.3393$	&	\textit{f}  &	$2.54^{+0.13}_{-0.08}$	&	$-10.50^{+0.11}_{-0.15}$	&	$3.21^{+0.16}_{-0.24}$	&	$-1.64\pm0.14$ \\
\textbf{060206	}	&	$	4.0559	$	&		&	$	2.61	^	{	+	0.16	}	_	{	-0.26	}	$	&	$	-10.32	^	{	+	0.07	}	_	{	-0.08	}	$	&	$	3.33	^	{	+	0.09	}	_	{	-0.11	}	$	&	$	-1.50	\pm	0.05	$	\\
\textbf{060210	}	&	$	3.9122	$	&		&	$	2.76	^	{	+	0.12	}	_	{	-0.17	}	$	&	$	-10.29	^	{	+	0.04	}	_	{	-0.04	}	$	&	$	3.39	^	{	+	0.06	}	_	{	-0.07	}	$	&	$	-1.59	\pm	0.06	$	\\
\textbf{060306	}	&	$	1.5591	$	&		&	$	2.25	^	{	+	0.15	}	_	{	-0.24	}	$	&	$	-10.35	^	{	+	0.04	}	_	{	-0.04	}	$	&	$	3.10	^	{	+	0.07	}	_	{	-0.09	}	$	&	$	-1.23	\pm	0.05	$	\\
\textbf{060319}&	$1.172$		&	\textit{b}  &	$1.79^{+0.11}_{-0.15}$	&	$-10.01^{+0.09}_{-0.12}$	&	$2.41^{+0.12}_{-0.17}$	&	$-1.03\pm0.02$ \\
\textbf{060502A}&	$1.51$		&	&	$2.67^{+0.18}_{-0.31}$	&	$-10.69^{+0.04}_{-0.04}$	&	$3.61^{+0.08}_{-0.10}$	&	$-1.35\pm0.06$ \\
\textbf{060512	}	&	$	2.1	$	&	\textit{e}	&	$	1.85	^	{	+	0.25	}	_	{	-0.19	}	$	&	$	-10.90	^	{	+	0.20	}	_	{	-0.40	}	$	&	$	2.98	^	{	+	0.24	}	_	{	-0.56	}	$	&	$	-1.45	\pm	0.17	$	\\
\textbf{060522}&	$5.11$		&	&	$2.69^{+0.20}_{-0.07}$	&	$-9.71^{+0.08}_{-0.10}$		&	$1.85^{+0.13}_{-0.19}$	&	$-1.57\pm0.10$ \\
\textbf{060604}	&	$2.1357$	&	&	$2.10^{+0.05}_{-0.06}$	&	$-11.13^{+0.06}_{-0.07}$	&	$3.62^{+0.11}_{-0.14}$	&	$-1.39\pm0.09$ \\
\textbf{060605}&	$3.773$		&	&	$2.69^{+0.18}_{-0.33}$	&	$-10.61^{+0.04}_{-0.04}$	&	$3.93^{+0.16}_{-0.25}$	&	$-5.82\pm1.79$ \\
\textbf{060607A	}	&	$	3.0749	$	&		&	$	2.68	^	{	+	0.10	}	_	{	-0.12	}	$	&	$	-9.74	^	{	+	0.04	}	_	{	-0.05	}	$	&	$	3.05	^	{	+	0.10	}	_	{	-0.14	}	$	&	$	-2.42	\pm	0.21	$	\\
\textbf{060708}&	$1.92$		&	&	$2.47^{+0.32}_{-0.06}$	&	$-10.36^{+0.08}_{-0.10}$	&	$2.76^{+0.10}_{-0.12}$	&	$-1.27\pm0.04$ \\
\textbf{060714}&	$2.7108$	&	\textit{g}  &	$2.37^{+0.16}_{-0.27}$	&	$-10.40^{+0.04}_{-0.04}$	&	$3.01^{+0.07}_{-0.09}$	&	$-1.47\pm0.06$ \\
\textbf{060904A	}	&	$	2.55	$	&		&	$	2.76	^	{	+	0.08	}	_	{	-0.09	}	$	&	$	-10.56	^	{	+	0.06	}	_	{	-0.08	}	$	&	$	3.01	^	{	+	0.16	}	_	{	-0.26	}	$	&	$	-1.37	\pm	0.11	$	\\
\textbf{061126	}	&	$	1.159	$	&		&	$	3.13	^	{	+	0.12	}	_	{	-0.16	}	$	&	$	-8.50	^	{	+	0.10	}	_	{	-0.14	}	$	&	$	2.19	^	{	+	0.10	}	_	{	-0.12	}	$	&	$	-1.37	\pm	0.02	$	\\
\textbf{061222A	}	&	$	2.088	$	&	\textit{e}	&	$	2.94	^	{	+	0.07	}	_	{	-0.08	}	$	&	$	-9.49	^	{	+	0.03	}	_	{	-0.03	}	$	&	$	3.00	^	{	+	0.04	}	_	{	-0.04	}	$	&	$	-1.38	\pm	0.02	$	\\
\textbf{070125	}	&	$	1.547	$	&		&	$	2.97	^	{	+	0.06	}	_	{	-0.07	}	$	&	$	-11.11	^	{	+	0.08	}	_	{	-0.09	}	$	&	$	4.62	^	{	+	0.12	}	_	{	-0.17	}	$	&	$	-2.65	\pm	0.33	$	\\
\textbf{070328	}	&	$	2.0627	$	&	\textit{e}	&	$	3.18	^	{	+	0.13	}	_	{	-0.12	}	$	&	$	-8.37	^	{	+	0.01	}	_	{	-0.01	}	$	&	$	2.15	^	{	+	0.02	}	_	{	-0.02	}	$	&	$	-1.61	\pm	0.02	$	\\
\textbf{070508	}	&	$	0.82	$	&		&	$	2.53	^	{	+	0.02	}	_	{	-0.02	}	$	&	$	-8.52	^	{	+	0.01	}	_	{	-0.01	}	$	&	$	2.42	^	{	+	0.02	}	_	{	-0.02	}	$	&	$	-1.46	\pm	0.02	$	\\
\textbf{070521	}	&	$	2.0865	$	&	\textit{e}	&	$	2.84	^	{	+	0.05	}	_	{	-0.04	}	$	&	$	-9.49	^	{	+	0.04	}	_	{	-0.05	}	$	&	$	2.70	^	{	+	0.07	}	_	{	-0.09	}	$	&	$	-1.55	\pm	0.09	$	\\
\textbf{071003	}	&	$	1.6044	$	&		&	$	3.31	^	{	+	0.04	}	_	{	-0.04	}	$	&	$	-10.53	^	{	+	0.13	}	_	{	-0.18	}	$	&	$	3.94	^	{	+	0.15	}	_	{	-0.22	}	$	&	$	-2.18	\pm	0.19	$	\\
\textbf{071031	}	&	$	2.6918	$	&	\textit{g}	&	$	1.95	^	{	+	0.11	}	_	{	-0.15	}	$	&	$	-11.61	^	{	+	0.13	}	_	{	-0.20	}	$	&	$	3.67	^	{	+	0.22	}	_	{	-0.46	}	$	&	$	-1.68	\pm	0.33	$	\\
\textbf{071117	}	&	$	1.331	$	&		&	$	2.05	^	{	+	0.18	}	_	{	-0.30	}	$	&	$	-10.64	^	{	+	0.11	}	_	{	-0.15	}	$	&	$	3.40	^	{	+	0.20	}	_	{	-0.38	}	$	&	$	-1.62	\pm	0.29	$	\\
\textbf{080319C	}	&	$	1.95	$	&		&	$	2.96	^	{	+	0.11	}	_	{	-0.16	}	$	&	$	-8.61	^	{	+	0.04	}	_	{	-0.04	}	$	&	$	2.68	^	{	+	0.07	}	_	{	-0.08	}	$	&	$	-1.68	\pm	0.06	$	\\
\textbf{080411	}	&	$	1.0301	$	&		&	$	2.72	^	{	+	0.06	}	_	{	-0.05	}	$	&	$	-9.19	^	{	+	0.04	}	_	{	-0.05	}	$	&	$	3.30	^	{	+	0.04	}	_	{	-0.05	}	$	&	$	-1.41	\pm	0.01	$	\\
\textbf{081028	}	&	$	3.038	$	&		&	$	2.37	^	{	+	0.15	}	_	{	-0.22	}	$	&	$	-12.01	^	{	+	0.19	}	_	{	-0.35	}	$	&	$	4.33	^	{	+	0.26	}	_	{	-0.79	}	$	&	$	-1.83	\pm	0.49	$	\\
\textbf{081109	}	&	$	0.9787	$	&		&	$	2.68	^	{	+	0.10	}	_	{	-0.12	}	$	&	$	-9.31	^	{	+	0.09	}	_	{	-0.12	}	$	&	$	2.27	^	{	+	0.11	}	_	{	-0.15	}	$	&	$	-1.17	\pm	0.03	$	\\
\textbf{081121	}	&	$	2.512	$	&		&	$	2.78	^	{	+	0.04	}	_	{	-0.05	}	$	&	$	-8.88	^	{	+	0.16	}	_	{	-0.24	}	$	&	$	2.41	^	{	+	0.14	}	_	{	-0.20	}	$	&	$	-1.50	\pm	0.03	$	\\
\textbf{081203A	}	&	$	2.05	$	&	\textit{e}	&	$	3.19	^	{	+	0.25	}	_	{	-0.15	}	$	&	$	-9.15	^	{	+	0.03	}	_	{	-0.03	}	$	&	$	2.52	^	{	+	0.05	}	_	{	-0.05	}	$	&	$	-1.85	\pm	0.06	$	\\
\textbf{081221	}	&	$	2.26	$	&	\textit{f}	&	$	2.45	^	{	+	0.04	}	_	{	-0.05	}	$	&	$	-8.47	^	{	+	0.15	}	_	{	-0.24	}	$	&	$	1.77	^	{	+	0.17	}	_	{	-0.29	}	$	&	$	-1.32	\pm	0.05	$	\\
\textbf{090205	}	&	$	4.6497	$	&		&	$	2.33	^	{	+	0.13	}	_	{	-0.18	}	$	&	$	-10.97	^	{	+	0.03	}	_	{	-0.04	}	$	&	$	3.31	^	{	+	0.10	}	_	{	-0.13	}	$	&	$	-2.63	\pm	0.33	$	\\
\textbf{090429B}&	$9.4$		&	&	$2.64^{+0.05}_{-0.06}$	&	$-10.89^{+0.06}_{-0.07}$	&	$2.51^{+0.15}_{-0.23}$	&	$-1.46\pm0.20$ \\
\textbf{090715B}&	$3.0$		&	&	$2.73^{+0.12}_{-0.17}$	&	$-10.11^{+0.20}_{-0.40}$	&	$2.56^{+0.21}_{-0.40}$	&	$-1.33\pm0.05$ \\
\textbf{090926A	}	&	$	2.1062	$	&	\textit{e}	&	$	2.95	^	{	+	0.04	}	_	{	-0.05	}	$	&	$	-10.47	^	{	+	0.26	}	_	{	-0.77	}	$	&	$	3.71	^	{	+	0.24	}	_	{	-0.61	}	$	&	$	-1.66	\pm	0.12	$	\\
\textbf{091003	}	&	$	0.8969	$	&		&	$	2.91	^	{	+	0.08	}	_	{	-0.09	}	$	&	$	-10.69	^	{	+	0.25	}	_	{	-0.62	}	$	&	$	4.06	^	{	+	0.25	}	_	{	-0.66	}	$	&	$	-1.56	\pm	0.14	$	\\
\textbf{091208B	}	&	$	1.063	$	&		&	$	2.39	^	{	+	0.04	}	_	{	-0.05	}	$	&	$	-9.63	^	{	+	0.05	}	_	{	-0.05	}	$	&	$	2.49	^	{	+	0.07	}	_	{	-0.09	}	$	&	$	-1.14	\pm	0.03	$	\\
\textbf{100728A	}	&	$	1.567	$	&		&	$	2.92	^	{	+	0.04	}	_	{	-0.05	}	$	&	$	-8.96	^	{	+	0.04	}	_	{	-0.04	}	$	&	$	2.78	^	{	+	0.05	}	_	{	-0.06	}	$	&	$	-1.56	\pm	0.04	$	\\
\textbf{100728B	}	&	$	2.106	$	&	\textit{e}	&	$	2.51	^	{	+	0.06	}	_	{	-0.07	}	$	&	$	-9.34	^	{	+	0.11	}	_	{	-0.14	}	$	&	$	1.58	^	{	+	0.17	}	_	{	-0.29	}	$	&	$	-1.17	\pm	0.07	$	\\
\textbf{101213A	}	&	$	0.414	$	&		&	$	2.64	^	{	+	0.15	}	_	{	-0.23	}	$	&	$	-10.23	^	{	+	0.05	}	_	{	-0.06	}	$	&	$	4.04	^	{	+	0.11	}	_	{	-0.14	}	$	&	$	-1.47	\pm	0.11	$	\\
\textbf{101219B	}	&	$	0.5519	$	&	\textit{a}	&	$	2.03	^	{	+	0.05	}	_	{	-0.05	}	$	&	$	-11.94	^	{	+	0.08	}	_	{	-0.09	}	$	&	$	4.48	^	{	+	0.19	}	_	{	-0.34	}	$	&	$	-1.06	\pm	0.17	$	\\
\textbf{110106B	}	&	$	0.618	$	&		&	$	2.29	^	{	+	0.03	}	_	{	-0.02	}	$	&	$	-10.23	^	{	+	0.04	}	_	{	-0.04	}	$	&	$	3.41	^	{	+	0.08	}	_	{	-0.09	}	$	&	$	-1.39	\pm	0.06	$	\\
\textbf{110205A}&	$2.22$		&	\textit{f}  &	$2.85^{+0.12}_{-0.18}$	&	$-9.24^{+0.24}_{-0.55}$		&	$2.43^{+0.19}_{-0.36}$	&	$-1.67\pm0.06$ \\
\textbf{110503A	}	&	$	1.613	$	&		&	$	2.74	^	{	+	0.04	}	_	{	-0.05	}	$	&	$	-8.38	^	{	+	0.05	}	_	{	-0.06	}	$	&	$	1.50	^	{	+	0.06	}	_	{	-0.07	}	$	&	$	-1.18	\pm	0.01	$	\\
\textbf{110715A	}	&	$	0.82	$	&		&	$	2.34	^	{	+	0.04	}	_	{	-0.05	}	$	&	$	-8.57	^	{	+	0.03	}	_	{	-0.03	}	$	&	$	2.10	^	{	+	0.05	}	_	{	-0.06	}	$	&	$	-1.23	\pm	0.03	$	\\
\textbf{110801A	}	&	$	1.858	$	&		&	$	2.60	^	{	+	0.13	}	_	{	-0.19	}	$	&	$	-10.41	^	{	+	0.06	}	_	{	-0.06	}	$	&	$	3.62	^	{	+	0.10	}	_	{	-0.12	}	$	&	$	-2.38	\pm	0.24	$	\\
\textbf{111008A	}	&	$	4.9898	$	&		&	$	2.95	^	{	+	0.10	}	_	{	-0.14	}	$	&	$	-10.19	^	{	+	0.03	}	_	{	-0.04	}	$	&	$	3.14	^	{	+	0.05	}	_	{	-0.05	}	$	&	$	-1.72	\pm	0.05	$	\\
\textbf{111209A	}	&	$	0.677	$	&		&	$	2.72	^	{	+	0.07	}	_	{	-0.08	}	$	&	$	-10.96	^	{	+	0.19	}	_	{	-0.34	}	$	&	$	4.32	^	{	+	0.23	}	_	{	-0.54	}	$	&	$	-1.44	\pm	0.16	$	\\
\textbf{120422A	}	&	$	0.283	$	&		&	$	1.52	^	{	+	0.34	}	_	{	-0.34	}	$	&	$	-12.35	^	{	+	0.06	}	_	{	-0.07	}	$	&	$	4.89	^	{	+	0.23	}	_	{	-0.51	}	$	&	$	-1.05	\pm	0.26	$	\\
\textbf{120712A	}	&	$	4.1745	$	&		&	$	2.81	^	{	+	0.08	}	_	{	-0.10	}	$	&	$	-9.70	^	{	+	0.05	}	_	{	-0.06	}	$	&	$	1.97	^	{	+	0.10	}	_	{	-0.13	}	$	&	$	-1.33	\pm	0.07	$	\\
\textbf{120729A}&	$0.8$		&	&	$2.32^{+0.66}_{-0.16}$	&	$-9.32^{+0.11}_{-0.15}$		&	$3.13^{+0.06}_{-0.07}$	&	$-2.63\pm0.15$ \\
\textbf{120909A	}	&	$	3.93	$	&		&	$	3.11	^	{	+	0.14	}	_	{	-0.21	}	$	&	$	-10.27	^	{	+	0.09	}	_	{	-0.11	}	$	&	$	3.09	^	{	+	0.13	}	_	{	-0.18	}	$	&	$	-1.46	\pm	0.12	$	\\
\textbf{120923A	}	&	$	8.5	$	&		&	$	2.58	^	{	+	0.09	}	_	{	-0.12	}	$	&	$	-11.34	^	{	+	0.10	}	_	{	-0.13	}	$	&	$	2.00	^	{	+	0.19	}	_	{	-0.36	}	$	&	$	-1.61	\pm	0.21	$	\\
\textbf{121027A}&	$1.773$		&	&	$2.12^{+0.21}_{-0.19}$	&	$-11.01^{+0.06}_{-0.07}$	&	$4.48^{+0.11}_{-0.15}$	&	$-1.45\pm0.11$ \\
\textbf{130514A	}	&	$	3.6	$	&		&	$	2.70	^	{	+	0.11	}	_	{	-0.15	}	$	&	$	-10.68	^	{	+	0.10	}	_	{	-0.12	}	$	&	$	3.31	^	{	+	0.17	}	_	{	-0.28	}	$	&	$	-1.40	\pm	0.16	$	\\
\textbf{130528A	}	&	$	1.25	$	&	\textit{b}	&	$	2.47	^	{	+	0.02	}	_	{	-0.02	}	$	&	$	-9.91	^	{	+	0.06	}	_	{	-0.08	}	$	&	$	2.76	^	{	+	0.10	}	_	{	-0.13	}	$	&	$	-1.11	\pm	0.05	$	\\
\textbf{130606A	}	&	$	5.913	$	&		&	$	3.31	^	{	+	0.12	}	_	{	-0.08	}	$	&	$	-10.92	^	{	+	0.05	}	_	{	-0.06	}	$	&	$	3.49	^	{	+	0.11	}	_	{	-0.15	}	$	&	$	-2.49	\pm	0.33	$	\\
\textbf{130907A	}	&	$	1.238	$	&	\textit{b}	&	$	2.94	^	{	+	0.01	}	_	{	-0.01	}	$	&	$	-7.72	^	{	+	0.01	}	_	{	-0.01	}	$	&	$	2.53	^	{	+	0.01	}	_	{	-0.01	}	$	&	$	-1.66	\pm	0.01	$	\\
\textbf{130925A}&	$0.347$		&	&	$2.39^{+0.02}_{-0.02}$	&	$-10.33^{+0.03}_{-0.03}$	&	$4.59^{+0.04}_{-0.05}$	&	$-1.31\pm0.03$ \\
\textbf{131108A	}	&	$	2.4	$	&		&	$	3.06	^	{	+	0.01	}	_	{	-0.01	}	$	&	$	-10.61	^	{	+	0.08	}	_	{	-0.10	}	$	&	$	3.44	^	{	+	0.11	}	_	{	-0.15	}	$	&	$	-1.89	\pm	0.15	$	\\
\textbf{140304A}&	$5.283$		&	&	$3.07^{+0.08}_{-0.09}$	&	$-9.21^{+0.05}_{-0.06}$		&	$2.48^{+0.12}_{-0.16}$	&	$-1.93\pm0.22$ \\
\textbf{140419A	}	&	$	3.956	$	&		&	$	3.16	^	{	+	0.11	}	_	{	-0.15	}	$	&	$	-9.27	^	{	+	0.03	}	_	{	-0.03	}	$	&	$	2.39	^	{	+	0.04	}	_	{	-0.05	}	$	&	$	-1.43	\pm	0.03	$	\\
\textbf{140423A	}	&	$	3.26	$	&		&	$	2.73	^	{	+	0.03	}	_	{	-0.03	}	$	&	$	-10.25	^	{	+	0.07	}	_	{	-0.09	}	$	&	$	2.97	^	{	+	0.11	}	_	{	-0.15	}	$	&	$	-1.45	\pm	0.09	$	\\
\textbf{140506A	}	&	$	0.889	$	&		&	$	2.43	^	{	+	0.10	}	_	{	-0.13	}	$	&	$	-9.59	^	{	+	0.07	}	_	{	-0.08	}	$	&	$	2.79	^	{	+	0.09	}	_	{	-0.11	}	$	&	$	-1.13	\pm	0.03	$	\\
\textbf{140512A	}	&	$	0.725	$	&		&	$	3.01	^	{	+	0.06	}	_	{	-0.07	}	$	&	$	-8.84	^	{	+	0.02	}	_	{	-0.02	}	$	&	$	2.80	^	{	+	0.04	}	_	{	-0.04	}	$	&	$	-1.29	\pm	0.02	$	\\
\textbf{140518A}&	$4.707$		&	&	$2.40^{+0.07}_{-0.08}$	&	$-10.61^{+0.05}_{-0.06}$	&	$2.89^{+0.17}_{-0.28}$	&	$-2.19\pm0.56$ \\
\textbf{140629A	}	&	$	2.275	$	&	\textit{f}	&	$	2.45	^	{	+	0.08	}	_	{	-0.10	}	$	&	$	-9.53	^	{	+	0.02	}	_	{	-0.02	}	$	&	$	2.47	^	{	+	0.05	}	_	{	-0.06	}	$	&	$	-1.63	\pm	0.06	$	\\
\textbf{140703A}&	$3.14$		&	&	$2.86^{+0.03}_{-0.04}$	&	$-9.83^{+0.07}_{-0.08}$		&	$2.64^{+0.15}_{-0.22}$	&	$-1.73\pm0.27$ \\
\textbf{140907A}&	$1.21$		&	\textit{b}  &	$2.40^{+0.03}_{-0.03}$	&	$-9.05^{+0.15}_{-0.22}$		&	$1.64^{+0.19}_{-0.34}$	&	$-1.07\pm0.04$ \\
\textbf{141220A	}	&	$	1.3195	$	&		&	$	2.62	^	{	+	0.02	}	_	{	-0.02	}	$	&	$	-8.54	^	{	+	0.19	}	_	{	-0.34	}	$	&	$	1.34	^	{	+	0.21	}	_	{	-0.41	}	$	&	$	-1.37	\pm	0.07	$	\\
\textbf{141221A	}	&	$	1.452	$	&	\textit{c}	&	$	2.57	^	{	+	0.07	}	_	{	-0.09	}	$	&	$	-9.58	^	{	+	0.12	}	_	{	-0.17	}	$	&	$	2.24	^	{	+	0.15	}	_	{	-0.23	}	$	&	$	-1.18	\pm	0.06	$	\\
\textbf{150120B}&	$3.5$		&	&	$2.77^{+0.33}_{-0.21}$	&	$-10.19^{+0.12}_{-0.16}$	&	$2.21^{+0.18}_{-0.31}$	&	$-1.32\pm0.13$ \\
\textbf{150206A	}	&	$	2.087	$	&	\textit{e}	&	$	2.85	^	{	+	0.07	}	_	{	-0.07	}	$	&	$	-10.89	^	{	+	0.06	}	_	{	-0.07	}	$	&	$	4.34	^	{	+	0.09	}	_	{	-0.12	}	$	&	$	-1.87	\pm	0.12	$	\\
\textbf{150314A	}	&	$	1.758	$	&	\textit{d}	&	$	2.93	^	{	+	0.01	}	_	{	-0.01	}	$	&	$	-8.03	^	{	+	0.02	}	_	{	-0.02	}	$	&	$	1.73	^	{	+	0.02	}	_	{	-0.02	}	$	&	$	-1.38	\pm	0.01	$	\\
\textbf{150323A}&	$0.593$		&	&	$1.99^{+0.06}_{-0.07}$	&	$-10.98^{+0.07}_{-0.08}$	&	$3.80^{+0.16}_{-0.26}$	&	$-1.48\pm0.19$ \\
\textbf{150403A	}	&	$	2.06	$	&	\textit{e}	&	$	3.06	^	{	+	0.07	}	_	{	-0.06	}	$	&	$	-8.32	^	{	+	0.01	}	_	{	-0.01	}	$	&	$	2.42	^	{	+	0.01	}	_	{	-0.01	}	$	&	$	-1.40	\pm	0.01	$	\\
\textbf{150821A}&	$0.755$		&	&	$2.75^{+0.02}_{-0.02}$	&	$-10.00^{+0.23}_{-0.54}$	&	$3.28^{+0.30}_{-1.86}$	&	$-1.27\pm0.18$ \\
\textbf{151021A	}	&	$	2.33	$	&	\textit{f}	&	$	2.75	^	{	+	0.06	}	_	{	-0.05	}	$	&	$	-9.19	^	{	+	0.14	}	_	{	-0.20	}	$	&	$	2.36	^	{	+	0.14	}	_	{	-0.21	}	$	&	$	-1.41	\pm	0.05	$	\\
\textbf{151027A	}	&	$	0.81	$	&		&	$	2.79	^	{	+	0.07	}	_	{	-0.09	}	$	&	$	-9.17	^	{	+	0.01	}	_	{	-0.01	}	$	&	$	3.65	^	{	+	0.03	}	_	{	-0.04	}	$	&	$	-2.32	\pm	0.09	$	\\
\textbf{151111A	}	&	$	3.5	$	&		&	$	2.68	^	{	+	0.03	}	_	{	-0.03	}	$	&	$	-10.59	^	{	+	0.09	}	_	{	-0.12	}	$	&	$	2.75	^	{	+	0.16	}	_	{	-0.27	}	$	&	$	-1.63	\pm	0.21	$	\\
\textbf{160227A	}	&	$	2.38	$	&		&	$	2.35	^	{	+	0.10	}	_	{	-0.12	}	$	&	$	-11.42	^	{	+	0.04	}	_	{	-0.04	}	$	&	$	4.91	^	{	+	0.10	}	_	{	-0.13	}	$	&	$	-2.53	\pm	0.32	$	\\
\textbf{160509A	}	&	$	1.17	$	&	\textit{b}	&	$	2.90	^	{	+	0.01	}	_	{	-0.01	}	$	&	$	-10.03	^	{	+	0.03	}	_	{	-0.03	}	$	&	$	4.28	^	{	+	0.06	}	_	{	-0.07	}	$	&	$	-1.74	\pm	0.08	$	\\
\textbf{160625B}&	$1.406$		&	\textit{c}  &	$3.20^{+0.01}_{-0.01}$	&	$-9.09^{+0.11}_{-0.16}$		&	$3.25^{+0.11}_{-0.14}$	&	$-1.46\pm0.03$ \\
\textbf{160804A	}	&	$	0.736	$	&		&	$	2.11	^	{	+	0.02	}	_	{	-0.02	}	$	&	$	-11.07	^	{	+	0.05	}	_	{	-0.06	}	$	&	$	3.79	^	{	+	0.12	}	_	{	-0.17	}	$	&	$	-1.07	\pm	0.08	$	\\
\textbf{161017A	}	&	$	2.0127	$	&	\textit{e}	&	$	2.95	^	{	+	0.06	}	_	{	-0.07	}	$	&	$	-10.37	^	{	+	0.04	}	_	{	-0.04	}	$	&	$	3.69	^	{	+	0.06	}	_	{	-0.07	}	$	&	$	-2.12	\pm	0.11	$	\\
\textbf{161117A	}	&	$	1.549	$	&		&	$	2.24	^	{	+	0.02	}	_	{	-0.02	}	$	&	$	-10.26	^	{	+	0.05	}	_	{	-0.06	}	$	&	$	3.33	^	{	+	0.08	}	_	{	-0.10	}	$	&	$	-1.26	\pm	0.06	$	\\
\textbf{161219B	}	&	$	0.1475	$	&		&	$	2.02	^	{	+	0.09	}	_	{	-0.11	}	$	&	$	-9.63	^	{	+	0.02	}	_	{	-0.02	}	$	&	$	3.52	^	{	+	0.03	}	_	{	-0.04	}	$	&	$	-1.01	\pm	0.01	$	\\
\textbf{170113A}&	$1.968$		&	&	$2.34^{+0.16}_{-0.27}$	&	$-9.89^{+0.04}_{-0.04}$		&	$2.79^{+0.06}_{-0.07}$	&	$-1.36\pm0.05$ \\
\textbf{170202A}&	$3.645$		&	&	$3.06^{+0.22}_{-0.19}$	&	$-10.38^{+0.07}_{-0.08}$	&	$2.77^{+0.12}_{-0.17}$	&	$-1.08\pm0.09$ \\
\textbf{170405A}&	$3.51$		&	&	$3.20^{+0.01}_{-0.01}$	&	$-9.80^{+0.05}_{-0.06}$		&	$2.54^{+0.11}_{-0.15}$	&	$-1.53\pm0.14$ \\
\textbf{170604A}&	$1.329$		&	&	$2.71^{+0.12}_{-0.11}$	&	$-10.08^{+0.08}_{-0.09}$	&	$3.29^{+0.12}_{-0.16}$	&	$-1.54\pm0.10$ \\
\textbf{170607A}&	$0.557$		&	\textit{a}  &	$2.38^{+0.04}_{-0.04}$	&	$-10.50^{+0.03}_{-0.03}$	&	$4.06^{+0.06}_{-0.06}$	&	$-1.20\pm0.04$ \\
\textbf{170705A}&	$2.01$		&	\textit{e}  &	$2.67^{+0.08}_{-0.07}$	&	$-10.17^{+0.02}_{-0.03}$	&	$3.43^{+0.04}_{-0.04}$	&	$-1.27\pm0.02$ \\
\textbf{171010A}&	$0.3285$	&	&	$2.31^{+0.01}_{-0.01}$	&	$-10.30^{+0.07}_{-0.08}$	&	$4.59^{+0.10}_{-0.13}$	&	$-1.93\pm0.13$ \\
\textbf{171205A}&	$0.0368$	&	&	$2.10^{+0.28}_{-0.13}$	&	$-11.86^{+0.05}_{-0.06}$	&	$5.30^{+0.13}_{-0.19}$	&	$-1.00\pm0.09$ \\
\textbf{180205A}&	$1.409$		&	\textit{c}  &	$2.31^{+0.07}_{-0.08}$	&	$-9.63^{+0.19}_{-0.33}$		&	$2.03^{+0.22}_{-0.49}$	&	$-1.08\pm0.06$ \\
\textbf{180325A}&	$2.248$		&	\textit{f}  &	$2.30^{+0.07}_{-0.06}$	&	$-9.17^{+0.03}_{-0.03}$		&	$3.11^{+0.05}_{-0.05}$	&	$-3.02\pm0.17$ \\
\textbf{180329B}&	$1.998$		&	\textit{e}  &	$2.16^{+0.07}_{-0.09}$	&	$-10.61^{+0.03}_{-0.03}$	&	$3.46^{+0.07}_{-0.09}$	&	$-1.81\pm0.15$ \\
\textbf{180620B}&	$1.1175$	&	\textit{b}  &	$2.50^{+0.03}_{-0.03}$	&	$-10.35^{+0.04}_{-0.04}$	&	$3.97^{+0.08}_{-0.10}$	&	$-1.48\pm0.09$ \\
\textbf{180720B}&	$0.654$		&	&	$3.02^{+0.01}_{-0.01}$	&	$-8.21^{+0.01}_{-0.01}$		&	$3.06^{+0.02}_{-0.02}$	&	$-1.47\pm0.01$ \\
\textbf{180728A}&	$0.117$		&	&	$1.95^{+0.01}_{-0.01}$	&	$-8.68^{+0.01}_{-0.01}$		&	$3.33^{+0.02}_{-0.02}$	&	$-1.30\pm0.01$ \\
\textbf{181010A}&	$1.39$		&	&	$2.83^{+0.11}_{-0.15}$	&	$-9.63^{+0.04}_{-0.05}$		&	$2.48^{+0.07}_{-0.08}$	&	$-1.25\pm0.04$ \\
\textbf{181020A}&	$2.938$		&	&	$3.16^{+0.02}_{-0.02}$	&	$-8.51^{+0.03}_{-0.03}$		&	$1.91^{+0.03}_{-0.04}$	&	$-1.69\pm0.03$ \\
\textbf{181110A}&	$1.505$		&	\textit{c}  &	$2.08^{+0.11}_{-0.36}$	&	$-10.05^{+0.05}_{-0.06}$	&	$3.13^{+0.08}_{-0.10}$	&	$-1.67\pm0.12$ \\
\hline
\end{longtable}

\end{appendix}

\end{document}